\documentclass[aps,superscriptaddress,showpacs,floatfix,secnumarabic,pre]{revtex4}
\usepackage{bm}
\usepackage{ragged2e}
\usepackage{amsmath}
\usepackage{amssymb}
\usepackage{graphicx}
\usepackage{cancel}
\usepackage{tabls}

\newcommand{\vect}[1]{\bm{#1}}
\newcommand{\fvect}[1]{\hat{#1}}
\newcommand{\high}{+}
\newcommand{\low}{-}
\newcommand{\measure}[2]{\frac{d^d #1\ d #2}{(2\pi)^{d+1}}}
\newcommand{\fmeasure}[2]{d\hat{#1}\ }
\newcommand{\eq}[1]{equation (\ref{#1})}

\begin{document}

%-------------------------------------------------------------------------------------------------------
\title{Re-examination of the infra-red properties of randomly stirred hydrodynamics}
%-------------------------------------------------------------------------------------------------------

\author{A. Berera}
\email{ab@ph.ed.ac.uk}
\affiliation{
SUPA, School of Physics and Astronomy,
University of Edinburgh, Edinburgh EH9 3JZ, UK}
\author{S. R. Yoffe}
\email{sam.yoffe@ed.ac.uk}
\affiliation{
SUPA, School of Physics and Astronomy,
University of Edinburgh, Edinburgh EH9 3JZ, UK}

\begin{abstract}

Dynamic renormalization group (RG) methods were originally used by Forster, 
Nelson and Stephen (FNS) to study the 
large-scale behaviour of randomly-stirred, incompressible fluids governed by 
the Navier-Stokes equations. 
Similar calculations using a variety of methods 
have been performed since, but have led to a discrepancy in results. 
In this paper, we carefully re-examine in 
$d$-dimensions the approaches used to calculate the renormalized viscosity 
increment and, by including an additional constraint which is neglected in many procedures, 
conclude that the original result of FNS is correct. 
By explicitly using 
step functions to control the domain of integration, we calculate a 
non-zero correction caused by boundary terms which cannot be ignored. 
We then go on to analyze how the noise renormalization, absent in many 
approaches, contributes an ${\mathcal O}(k^2)$ correction to the force 
autocorrelation and show conditions for this to be taken as a
renormalization of the noise coefficient. Following this, we discuss the 
applicability of this RG procedure to the calculation of the inertial 
range properties of fluid turbulence.

\end{abstract}

%\date{\today}

\pacs{47.27.ef, 47.27.-i, 11.10.Gh}

\maketitle
\medskip
\begin{center}
\textit{In press Physical Review E, 2010.}
\end{center}

\section{Introduction}\label{sec:intro}
The large scale behaviour of randomly stirred fluids was originally studied by Forster, Nelson and Stephen (FNS) \cite{FNS:1976-letter,FNS:1977-full}. They used a dynamic renormalization procedure to explore the effects of the progressive removal of small (length) scales in a perturbative model under several types of forcing. As they note, their study is only valid at the smallest momentum scales, and as such the study is well below the momentum scale of the inertial range \cite{McComb:2006-ngpt}. Later, the procedure used by FNS was extended by Yakhot \& Orszag (YO) \cite{YO:1986-RG_letter, YO:1986-RG_full} to a more general forcing spectrum (of which the studies of FNS were special cases) and used to calculate the energy spectrum and a value for the Kolmogorov constant in the inertial region.
%--- @Arjun: Is this sentence needed?
While their arguments allowing them to calculate inertial range properties are contested \cite{Teodorovich:1994-YO_theory,mccomb:1990-book, Eyink:1994-RG_method_stat_hyd}, these issues are not the main focus of this paper. Instead, we will concentrate on another disagreement related to the results for the renormalized viscosity and noise.
%---

In the papers of FNS \cite{FNS:1977-full} and YO \cite{YO:1986-RG_full}, the authors calculate the viscosity increment, quantifying the effect of the removed subgrid scales on the super-grid scales. They find the prefactor $\tilde{A}_d(\epsilon) = (d^2 - d - \epsilon)/{2d(d+2)}$ (from \cite{YO:1986-RG_full}, with FNS in agreement for their specific cases of study). The disagreement is centred around the use of a certain change of variables employed by FNS and YO. This substitution has been highlighted as a cause for concern (for example, 
\cite{Wang_Wu:1993-mod_YO, Teodorovich:1994-YO_theory}), since na\"{\i}vely the symmetric domain of integration appears to be shifted, violating conditions for the identities used to be valid. Using methods that do not introduce any substitution, again for a general forcing spectrum, 
Wang \& Wu (WW) \cite{Wang_Wu:1993-mod_YO} and 
Teodorovich \cite{Teodorovich:1994-YO_theory} arrive at a different, 
incompatible result for the viscosity increment. Instead, they find 
the prefactor $\tilde{A}^\star_d = (d - 1)/{2(d+2)} = \tilde{A}_d(0)$. 
This $\epsilon$-free result is also used in the more field-theoretic work of 
Adzhemyan \textit{et al.} \cite{AAV:1999-book}. Later, 
Nandy \cite{Nandy:1997-symm} attempted to determine which of the results 
was correct using a ``symmetrization argument'' and agreed with the 
original (general forcing) result by YO.

The method used by FNS and YO has found wide-ranging application, for 
example in soft matter systems, such as the KPZ and Burgers' equations  
\cite{KPZ:1986-scaling,Burgers:1948-eq,MHKZ:1989-correlated,Frey_Tauber:1994-two-loop}, and the coupled equations of magnetohydrodynamics 
\cite{FSP:1982-ir_mhd}. 
Given the extensive use of this approach, it is unsatisfactory to have any 
lingering disagreement on the basic methodology. The aim of this paper is, 
therefore, to settle this dispute once and for all. There cannot be two 
different results for the same quantity. We will show that an extra 
constraint mentioned by FNS causes the elimination band not to be shifted, 
and that for substitution-free methods there are neglected boundary terms.
These are evaluated and shown to compensate exactly the difference between 
$\tilde{A}^\star_d$ and $\tilde{A}_d$. We then show how correct treatment 
does not require a symmetrization to obtain the Yakhot-Orszag result.

In addition to renormalization of the viscosity, there is also renormalization
of the noise.  All treatments consider an input noise that is
Gaussian with the forcing spectrum parametrized as $W_0 k^{-y}$,
where $k$ is the wavenumber associated with the force.
At one-loop order each of the two vertices will have a factor of the inflowing
momentum,  thus leading to a $k^2$ contribution to the
forcing spectrum.  Both FNS and YO acknowledge this $k^2$ correction.
In FNS they treat two specific cases, $y= -2$ and $y=0$.
In the former, they find a renormalization to $W_0$ whereas, in the
latter, they conclude that all higher order corrections are subleading.
YO restrict their analysis to $y > -2$ and once again conclude
all higher order corrections are subleading. We explicitly show how the
leading contribution will always go as $k^2$ and as such can only be taken
as an multiplicative renormalization for the case $y=-2$, as noted in
\cite{YOD:1987-analytic_theories}. We find the prefactor agrees with
$A_d(\epsilon)$ found by FNS and YO (with $y=-2$) and show it to be incompatible
with the $\epsilon$-independent $A_d^\star$.

Another author, Ronis \cite{Ronis:1987-field_theor_RG}, 
calculates the viscosity and noise
renormalization using a field-theoretic approach.
His analysis agrees with FNS and YO for $y=-2$, although
appears to be presented for general $y$. 
As we will argue in Sec. \ref{sec:renorm_cond}, this seems
unjustified as the noise is only renormalized for the case 
$y=-2$.

The paper is organised as follows. In Sec. \ref{sec:discuss}, we give a brief
discussion on the validity and limitation of this type of low-$k$
renormalization scheme for a fluid system. In Sec. \ref{sec:calculation} 
our calculation
for viscosity renormalization is done and then in 
Sec. \ref{sec:renorm_cond} for
noise renormalization, along with comparison with other analyses.
The results are 
summarised in table \ref{tbl:summary}.
Finally, in Sec. \ref{sec:conclusion}, we present our conclusions along with
a brief discussion of the relevance of this type
of renormalization scheme for calculating inertial range quantities.

\section{Discussion and relevance of approach}
\label{sec:discuss}

We start with a brief discussion on the region of validity of this method 
and its limitations. Turbulence is often viewed as an energy cascade, 
where energy enters large length scales in the production range and is 
progressively transferred to smaller and smaller scales, until viscous 
effects dominate and it is dissipated as heat. There must be a balance 
between the energy dissipated and the energy transferred through the 
intermediate scales, otherwise energy would build up and the turbulence 
would not remain statistically steady. Thus the dissipation rate, $\varepsilon$, controls how small the smallest length scales need to be to successfully remove the energy passed down, giving the Kolmogorov scale $\eta = \left(\nu^3/\varepsilon\right)^{1/4} = 1/k_d$. When the Reynolds number is sufficiently large, there exists a range of intermediate scales where the energy flux entering a particular length scale from ones larger than it is the same as that leaving it to smaller ones and is thus not dependent on the wavenumber. 
This is the inertial range.

\begin{figure}
 \centering
 \includegraphics[bb=99 220 487 445, width=0.65\textwidth]{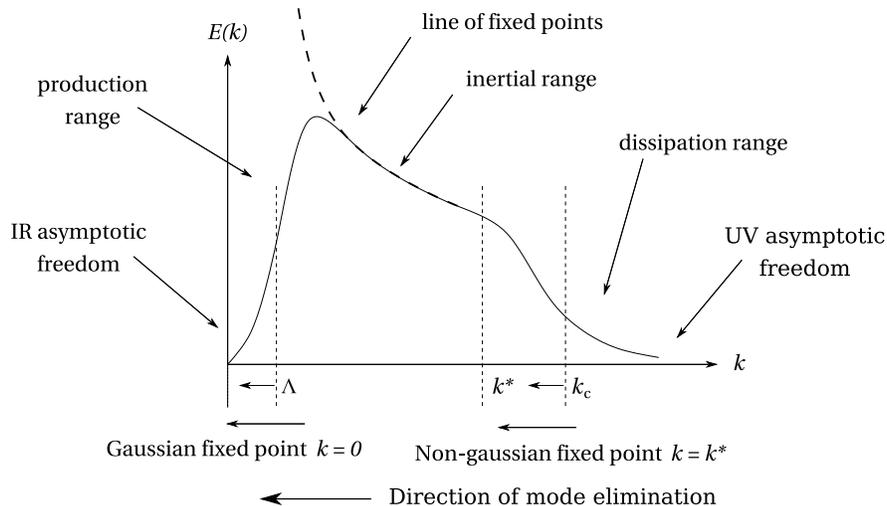}
 \caption{The energy spectrum for a turbulent flow. Renormalization from $\Lambda \ll k_d$ introduced by FNS in their IR study of randomly stirred flows takes you to the fixed point at $k = 0$. Iterative averaging from $k_c \simeq 0.1 k_d$ takes you to the non-Gaussian fixed point $k^*$, which marks the beginning of a line of fixed points along $k^{-5/3}$.}
 \label{fig:energy_spectrum}
\end{figure}

The energy spectrum and a summary of the various ranges of it are
presented in Fig. \ref{fig:energy_spectrum} (based on a similar
figure in \cite{McComb:2006-ngpt}).
In the RG approach, the smallest length scales (largest wavenumber scales
$k$) are removed and an effective 
theory is obtained from the remaining scales. There is high- and low-energy asymptotic freedom since the renormalized coupling becomes weak in both limits \cite{McComb:2006-ngpt}.
The dynamic RG method used by FNS introduces 
a momentum cutoff $\Lambda$ well below the dissipation 
momentum scale, below the 
inertial range even (see figure \ref{fig:energy_spectrum}), in the production 
range and removes momentum scales towards $k=0$. As such, this method can 
only ever account for the behaviour on the largest length scales. 
The production range is highly dependent on the method of energy input, and 
so it is obvious that the properties of the lowest modes will also share 
this dependence. Taking the forcing to be Gaussian then allows Gaussian 
perturbation theory to be used, since the lowest order is simply the 
response to this forcing. Since the inertial range is highly non-Gaussian,
we do not expect to study the inertial region with this analysis.

An alternative RG scheme called iterative averaging (McComb \cite{McComb:1982-reform, McComb:1990-ca, McComb:1992-two_field, McComb:2006-ngpt}) instead takes a cutoff $k_c \sim 0.1 k_d$ and removes successive shells of wavenumbers down to a non-Gaussian fixed point $k^*$, which marks the beginning of a line of fixed points following $k^{-5/3}$ through the inertial region (see figure \ref{fig:energy_spectrum}). The asymptotic nature of this method therefore cannot tell us anything about the forcing spectrum, and is only dependent on the rate at which energy is given to the system. No assumptions about Gaussian behaviour are made.

Using the energy spectrum (figure \ref{fig:energy_spectrum}), we see the 
location of the IR procedure of FNS/YO and how it is inapplicable for 
the calculation of inertial range statistics. Put simply, it does not 
have access to the inertial range, just as iterative averaging does not 
have access to the production range. This is discussed further in 
the conclusions, Sec. \ref{sec:conclusion}.

\section{Calculation}\label{sec:calculation}
The motion of an incompressible Newtonian fluid in $d$-spatial dimensions, subject to stochastic forcing, $\vect{f}$, is governed by the Navier-Stokes equation (NSE) which, in configuration space, is
\begin{equation}
 \label{eq:def:nse_conf}
 \frac{\partial u_\alpha(\vect{x},t)}{\partial t} + \Big(u_\beta(\vect{x},t) \frac{\partial}{\partial x_\beta}\Big) u_\alpha(\vect{x},t) = -\frac{1}{\rho} \frac{\partial p(\vect{x},t)}{\partial x_\alpha}  + \nu_0 \nabla^2 u_\alpha(\vect{x},t) + f_\alpha(\vect{x},t) \ ,
 \end{equation}
where $\vect{u}(\vect{x},t)$ is the velocity field, $p(\vect{x},t)$ is the pressure field, $\rho$ is the density of the fluid and $\nu_0$ is the kinematic viscosity. The index $\alpha \in \{ 1, \ldots, d \}$ and there is an implied summation over repeated indices. We consider an isotropic, homogeneous fluid and, using the Fourier transform defined by
\begin{equation}
 \label{transform_x}
 u_\alpha(\vect{x},t) = \int \frac{d^dk}{(2\pi)^d}\frac{d\omega}{(2\pi)} u_\alpha(\vect{k},\omega) e^{i\vect{k}\cdot\vect{x} + i\omega t}\ ,
\end{equation}
the NSE may be expressed in Fourier-space as
\begin{equation}
 \label{eq:def:nse_fourier}\begin{split}
 \left(i\omega + \nu_0 k^2\right) u_\alpha(\vect{k},\omega) = f_\alpha(\vect{k},\omega) + \lambda_0 M_{\alpha\beta\gamma}(\vect{k}) \int \measure{j}{\Omega} \int\measure{p}{\Omega'} u_\beta(\vect{j},\Omega) u_\gamma(\vect{p},\Omega') \\
(2\pi)^{d+1} \delta(\vect{j}+\vect{p}-\vect{k}) \delta(\Omega + \Omega' - \omega)
\end{split}
\end{equation}
\begin{equation}
 \label{eq:cont_fourier}
 k_\alpha u_\alpha(\vect{k},\omega) = 0\ ,
\end{equation}
where the incompressibility condition ($\vect{\nabla}\cdot\vect{u} = 0$) has been used to solve for the pressure field in terms of the velocity field. In \eq{eq:def:nse_fourier}, we have also introduced $\lambda_0$ ($=1$) as a book-keeping parameter to the non-linear term and the vertex and projection operators, respectively, are defined as
\begin{equation}
 \label{eq:def:vert_proj}
 \begin{split}
  M_{\alpha\beta\gamma}(\vect{k}) &= \frac{1}{2i}\Big[k_\beta P_{\alpha\gamma}(\vect{k}) + k_\gamma P_{\alpha\beta}(\vect{k}) \Big]\ , \\
  P_{\alpha\gamma}(\vect{k}) &= \delta_{\alpha\gamma} - \frac{k_\alpha k_\gamma}{k^2}\ ,
 \end{split}
\end{equation}
and contain the contribution from the pressure field. The integral over $\vect{p}$, $\Omega'$ could be trivially done 
to follow FNS \cite{FNS:1977-full}, YO \cite{YO:1986-RG_full} and Wang \& Wu
\cite{Wang_Wu:1993-mod_YO}; however, we leave it in for comparison 
with Nandy's calculation \cite{Nandy:1997-symm}. 
It is common to specify the forcing term through 
its autocorrelations
\begin{equation}
\label{eq:def:stirring_forces}
 \langle f_\alpha(\vect{k},\omega) f_\beta(\vect{k'},\omega') \rangle = 2 W(k) P_{\alpha\beta}(\vect{k}) (2\pi)^{d+1} \delta(\vect{k} + \vect{k'}) \delta(\omega + \omega')\ ,
\end{equation}
where $W(k) = W_0 k^{-y}$ is the forcing spectrum and the presence of the projection operator guarantees that the forcing is solenoidal (and hence maintains the incompressibility of the velocity field). Since the RHS is real and symmetric under $\vect{k}\to-\vect{k}$, the configuration-space correlation is also real.

Following FNS \cite{FNS:1977-full}, we impose a hard UV cut-off 
$\Lambda \ll k_d$, where $k_d$ is the dissipation wavenumber.
With this choice of cut-off the theory only 
accounts for the largest scale behaviour (and therefore should not reproduce 
results for inertial range turbulence). 
This cut-off was later relaxed to $\Lambda \sim {\mathcal O}(k_d)$ by YO,
although the rest of their renormalization procedure followed FNS.
The velocity field can then be decomposed into its high and low 
frequency modes, introducing a more compact notation as
($u_\alpha^{\low}$ and $u_\alpha^{\high}$ are often also
expressed as $u_\alpha^{<}$ and $u_\alpha^{>}$),
\begin{equation*}
\label{eq:def:decomposition}
u_\alpha(\fvect{k}) = \left\{ \begin{array}{c}
u^{\low}_\alpha(\fvect{k}) \qquad 0 < \lvert \vect{k}\rvert < e^{-\ell}\Lambda \\
\\
u^{\high}_\alpha(\fvect{k}) \qquad e^{-\ell}\Lambda < \lvert \vect{k}\rvert < \Lambda
\end{array} \right. \ ; \quad \ell > 0\ ,
\end{equation*}
with $\fvect{k} = (\vect{k},\omega)$ such that $\fmeasure{k}{\omega} = (2\pi)^{-(d+1)} d^dk\ d\omega$ and $\delta(\fvect{k}) = (2\pi)^{d+1} \delta(\vect{k})\delta(\omega)$, allowing the NSE to be rewritten for the component fields:
\begin{equation}
 \label{eq:nse_low}
\begin{split}
 \left(i\omega + \nu_0 k^2\right)u_\alpha^{\low}(\fvect{k}) = f_\alpha^\low(\fvect{k})
+ \lambda_0 M^\low_{\alpha\beta\gamma}(\vect{k}) \int\fmeasure{j}{\Omega} \int\fmeasure{p}{\Omega'} \Big[u_\beta^{\low}(\fvect{j}) u_\gamma^{\low}(\fvect{p}) + 2u_\beta^{\low}(\fvect{j}) u_\gamma^{\high}(\fvect{p}) + u_\beta^{\high}(\fvect{j}) u_\gamma^{\high}(\fvect{p}) \Big] \\
\delta(\fvect{j}+\fvect{p}-\fvect{k})\ ,
\end{split}
\end{equation}
and similarly for the high frequency modes, $u_\alpha^\high(\fvect{k})$. 
The filtered vertex operator $M^\low_{\alpha\beta\gamma}(\vect{k})$ is
understood to restrict $0 < k < e^{-\ell}\Lambda$ in the non-linear term.  
This will later lead to an additional constraint on the loop integral.
This constraint is neglected by many authors.

Together with a perturbation expansion and the zero-order propagator,
\begin{equation}
\begin{split}
 u_\alpha^{\high}(\fvect{k}) &= u_\alpha^{\high(0)}(\fvect{k}) + \lambda_0 u_\alpha^{\high(1)}(\fvect{k}) + \lambda_0^2 u_\alpha^{\high(2)}(\fvect{k}) + \ldots \\
 G_0(\fvect{k}) &= \frac{1}{i\omega + \nu_0 k^2}\ ,
\end{split}
\end{equation}
it is possible to solve for $u_\alpha^{\high(i)}(\fvect{k})$ in terms of $u_\alpha^{\high(0)}(\fvect{k}) = G_0(\fvect{k}) f_\alpha^\high(\fvect{k})$ using powers of $\lambda_0$, which may be substituted back into \eq{eq:nse_low}. Performing a filtered-averaging procedure $\langle\cdots\rangle_f$, under which:
\begin{enumerate}
 \item Low frequency components are statistically independent of high frequency components.
 \item Low frequency components invariant under averaging: $\langle f^{\low}\rangle_f \simeq f^{\low}$ and so $\langle u^{\low}\rangle_f = u^{\low} + {\cal{O}}(\lambda_0)$.\\
(This can be shown more rigorously using a \textit{conditional average} \cite{thesis:ahunter, McComb:1992-ca}, and discussed in \cite{Sukoriansky_Galperin_Staroselsky:2003-cross_term}.)
 \item Stirring forces are Gaussian with zero mean: $\langle f^{\high}\rangle_f = 0$, $\langle f^{\high} f^{\high} f^{\high}\rangle_f = 0$;
\end{enumerate}
and using \eq{eq:def:stirring_forces}, we obtain
\begin{equation}
 \label{eq:post_av}
 \left(i\omega + \nu_0 k^2 \right) u^{\low}_\alpha(\fvect{k}) = \Big[f^{\low}_\alpha(\fvect{k}) + \Delta f^{\low}_\alpha (\fvect{k})\Big] + \lambda_0 M^\low_{\alpha\beta\gamma}(\vect{k})\int^\low \fmeasure{j}{\Omega} \int^\low \fmeasure{p}{\Omega'} u_\beta^{\low}(\fvect{j}) u_\gamma^{\low}(\fvect{p}) \delta(\fvect{p}-\fvect{k}+\fvect{j}) + \Sigma^\low_{\alpha}(\fvect{k}) \ ,
\end{equation}
where
\begin{equation}
 \label{eq:sigma_alpha}
\begin{split}
 \Sigma^\low_\alpha(\fvect{k}) = 8 \lambda_0^2 M_{\alpha\beta\gamma}(\vect{k}) \theta^\low(\vect{k}) \int\fmeasure{j}{\Omega}\hspace{-1mm}\int\fmeasure{p}{\Omega'} M_{\gamma\mu\nu}(\vect{p}) G_0(\fvect{p}) \left\lvert G_0(\fvect{j}) \right\rvert^2 W(j) P_{\beta\nu}(\vect{j}) u_\mu(\fvect{p}+\fvect{j}) \\
\theta^{\low}(\vect{p}+\vect{j}) \theta^{\high}(\vect{j}) \theta^{\high}(\vect{p}) \delta(\fvect{p}-\fvect{k}+\fvect{j})\ ,
\end{split}
\end{equation}
and we have used $\theta$-functions,
\begin{eqnarray}
 \label{eq:def:theta}
 \theta^{\low}(\vect{k}) = \theta(\Lambda e^{-\ell}-\lvert\vect{k}\rvert) &=& \left\{ \begin{array}{ll}
1 & \lvert\vect{k}\rvert < \Lambda e^{-\ell} \\
1/2 & \lvert\vect{k}\rvert = \Lambda e^{-\ell} \\
0 & \textrm{otherwise}
\end{array} \right.\ \nonumber\\
\theta^\high(\vect{k}) = \theta(\lvert\vect{k}\rvert-\Lambda e^{-\ell}) \theta(\Lambda-\lvert\vect{k}\rvert) &=& \left\{ \begin{array}{ll}
1 & \Lambda e^{-\ell} < \lvert\vect{k}\rvert < \Lambda \\
1/2 & \lvert\vect{k}\rvert = \Lambda e^{-\ell} \\
0 & \textrm{otherwise}
\end{array} \right.\ ,
\end{eqnarray}
to explicitly control the shell of integration, so the momentum integrals are now $0 < \lvert \vect{\kappa} \rvert < \infty$. The induced random force,
\begin{equation}
\label{eq:induced_force}
\Delta f^\low_\alpha(\fvect{k}) = \lambda_0 M^\low_{\alpha\beta\gamma}(\vect{k})  \int \fmeasure{j}{\Omega} G_0(\fvect{j}) f_\beta(\fvect{j}) f_\gamma(\fvect{k}-\fvect{j}) G_0(\fvect{k}-\fvect{j}) \theta^\high(\vect{j}) \theta^\high(\vect{k}-\vect{j}) \ ,
\end{equation}
compensates for the effect of forcing on the eliminated modes -- see section \ref{sec:renorm_cond} for more details. Note that in \eq{eq:post_av} we have, following \cite{FNS:1977-full} and \cite{YO:1986-RG_full}, neglected the velocity triple non-linearity (and thus all higher non-linearities which are generated by it). Eyink \cite{Eyink:1994-RG_method_stat_hyd} showed that this operator is not irrelevant but marginal by power counting (see appendix \ref{app:operator_scaling}); however, as noted in \cite{ZhouMcCombVahala:1997-RG_in_turb}, this choice merely indicates the order of approximation and doesn't require justification. In any case, these higher-order operators become irrelevant as $k \rightarrow 0$ \cite{McComb:2006-ngpt}.

\begin{figure}[tb]
 \begin{minipage}[t]{\columnwidth}
  \centering
  \includegraphics[width=0.6\linewidth,bb=125 605 539 715]{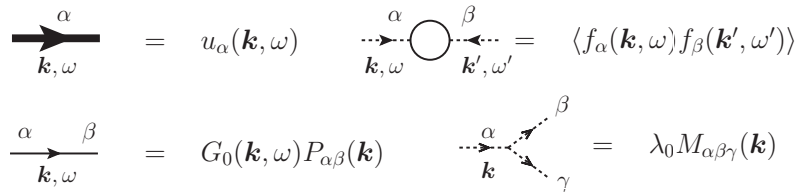} %bb=left bottom right top
  \caption{Feynman rules for the velocity, force autocorrelation, propagator and vertex. These modifications clarify those in figure 3 of \cite{FNS:1977-full}, since their vertex operator appears to be connected to the propagator. The dotted lines represent any combination of the solid lines, shown left.}
  \label{fig:rules}
 \end{minipage}
\end{figure}
\begin{figure}[tb]
 \begin{minipage}[t]{0.95\columnwidth}
  \centering
   \includegraphics[width=0.5\columnwidth,bb=71 634 569 720]{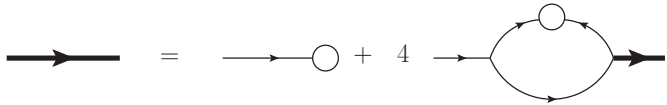}
  \caption{Renormalization of the velocity field giving the one-loop correction to the viscosity. See figure \ref{fig:rules} for Feynman rules. From the factor $4$ (see \cite{Wyld:1961-formulation}), $2$ comes from exchanging which leg from the left vertex connects to the noise correlation, and the other $2$ from the thick line instead being incident on the LHS.}
  \label{fig:v_graph}
 \end{minipage}
\end{figure}
Multiplying both sides of equation (\ref{eq:post_av}) by $G_0(\fvect{k})$ and neglecting the triple non-linearity, this expression can be found from the graph given in figure \ref{fig:v_graph} using the rules in figure \ref{fig:rules} and the form in equation (\ref{eq:sigma_alpha}). It should be noted that the symmetry factor of the graph is 4 (figure 9 in \cite{Wyld:1961-formulation}).

From \eq{eq:sigma_alpha}, we may perform the frequency integrals in either order to give
\begin{equation}
 \label{eq:all_agree}
 \Sigma^\low_\alpha = \frac{4\lambda_0^2}{\nu_0} M_{\alpha\beta\gamma}(\vect{k}) \theta^\low(\vect{k})\int \frac{d^dj}{(2\pi)^d} \frac{W(j)}{j^2} \int d^dp\ M_{\gamma\mu\nu}(\vect{p}) \frac{P_{\beta\nu}(\vect{j}) u^\low_\mu(\vect{p}+\vect{j},\omega)}{i\omega + \nu_0 j^2 + \nu_0 p^2}\ \theta^{\high}(\vect{j}) \theta^{\high}(\vect{p}) \delta(\vect{p}-\vect{k}+\vect{j}) \ ,
\end{equation}
which, along with the definition of $W(j)$ and $M_{\alpha\beta\gamma}(\vect{k}) = (2i)^{-1} P_{\alpha\beta\gamma}(\vect{k})$, may be compared to (2.10) in \cite{YO:1986-RG_full} or (4) in \cite{Wang_Wu:1993-mod_YO}.

\subsection{Analysis of the self-energy integral}\label{subsec:self_energy_analysis}
We begin this section with the motivation for calling this a self-energy integral. The term has been borrowed from high-energy physics, and it represents the field itself modifying the potential it experiences. In high-energy physics, the renormalized or dressed propagator may be written using the Dyson equation \cite{Wyld:1961-formulation} (equation (27)) as $G = G_0 + G_0 \Sigma G + \cdots$, where $\Sigma$ represents the self energy operator. In our case, we are instead writing $u_\alpha(k) = G_0(k) f_\alpha(k) + G_0(k) \Sigma_{\alpha\mu} u_\mu(k)$, where the structure $\Sigma_\alpha = \Sigma_{\alpha\mu} u_\mu$ can be seen from the graph in figure \ref{fig:v_graph} or equation (\ref{eq:all_agree}) once the integral over $\vect{p}$ has been trivially done.

As can be seen in \eq{eq:all_agree}, the constraints on the integral are provided by the product of $\theta^\high$ functions. We first show how this can be expanded before verifying that at ${\mathcal O}(k)$ the substitution causes two compensating corrections, and hence to ${\mathcal O}(k^2)$ there is no correction. 
Following this, corrections to the calculations by Wang \& Wu and Nandy 
are evaluated and their contribution to the final result accounted for.

We perform the integral over $d^dp$ so our product of $\theta^\high$ 
functions becomes $\theta^\high(\vect{j}) \theta^\high(\vect{k}-\vect{j})$. 
The second constraint, $\theta^\high(\vect{k}-\vect{j})$, is sometimes
ignored (see for example equation (4) in \cite{Wang_Wu:1993-mod_YO}) and this is
a source of error in these calculations. 
With the definition $\theta^\high(\vect{\kappa}) = \theta(\lvert\vect{\kappa}\rvert - D) \theta(\Lambda-\lvert\vect{\kappa}\rvert)$ from \eq{eq:def:theta} where $D = \Lambda e^{-\ell}$, we Taylor expand the latter about $\vect{k} = 0$ and our product becomes
\begin{equation}
 \label{eq:ww_theta_exp}
 \theta^\high(\vect{j})\theta^\high(\vect{k}-\vect{j}) = \theta^\high(\vect{j})\left(1 - \frac{\vect{k}\cdot\vect{j}}{\lvert\vect{j}\rvert} \Big[\theta(\Lambda-\lvert\vect{j}\rvert) \delta(\lvert\vect{j}\rvert - D) - \theta(\lvert\vect{j}\rvert-D) \delta(\Lambda-\lvert\vect{j}\rvert) \Big]\right) + {\mathcal O}(k^2)\ ,
\end{equation}
see appendix \ref{app:theta_exp} for details. We see that the additional constraint has introduced a first order correction to the constraint on $\lvert\vect{j}\rvert$. Further, the presence of the $\delta$-functions show that these contributions are evaluated on the boundaries. This correction is absent from the work of Wang \& Wu, Teodorovich and Nandy as they ignore this constraint. We shall see later that, from a diagrammatic point of view, this is equivalent to ensuring that all internal lines have momenta in the eliminated band.

\subsubsection{FNS and YO}
\begin{figure}
 \centering
 \includegraphics[width=0.85\textwidth,bb=113 543 308 625]{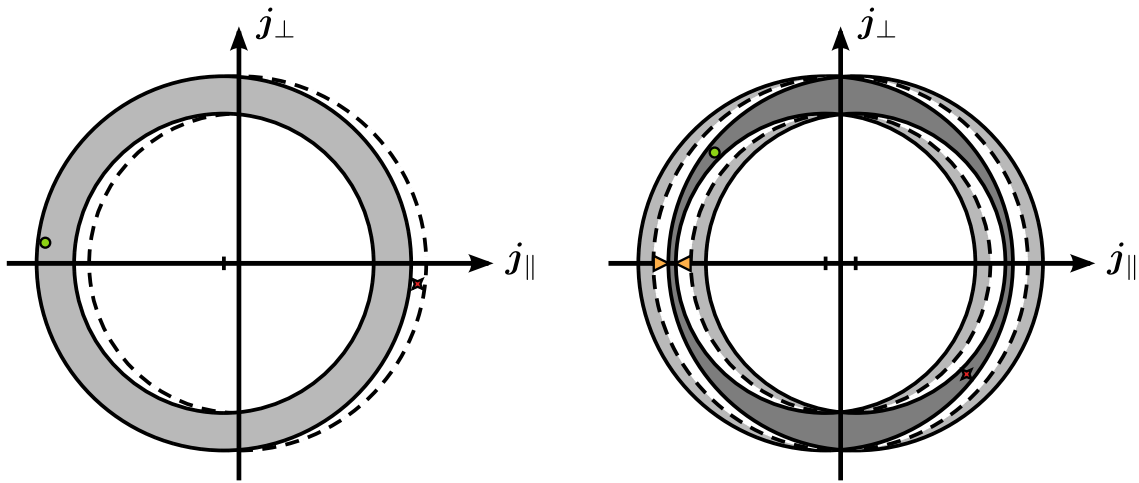}
 \caption{(Color online) The shifted shell of integration caused by the substitution 
used by FNS/YO. $\vect{j}_{\parallel}$ and $\vect{j}_{\perp}$ label the momenta in the directions parallel and perpendicular to that of the shift, $\hat{\vect{k}}$. (Left) The constraint  $\theta^{+}(\vect{j}+\tfrac{1}{2}\vect{k})$ only, as used by Wang \& Wu; (right) $\theta^{+}(\vect{j}+\tfrac{1}{2}\vect{k})\theta^{+}(\vect{j}-\tfrac{1}{2}\vect{k})$ as used by FNS. White ($\cdot - \cdot$) shows the position of the unshifted shell $\theta^{+}(\vect{j})$, with light grey shading (-\hspace{0.5mm}-\hspace{0.5mm}-) showing the shifted shell(s). Small tick marks indicate the centre for the shifted shell(s) at $\vect{j} = \mp\tfrac{1}{2}\vect{k}$. The dark grey highlights the overlap of two shifted shells. The green point (in the top-left quadrant) is a random momentum, $\vect{q}$, which lies within the resultant shell, while the red cross (lower-right quadrant) shows $-\vect{q}$. Clearly, in the left case the shell is not symmetric under $\vect{j} \to -\vect{j}$, and as such identities requiring a symmetric shell are invalid. By correctly accounting for the additional momentum constraint, we are led instead to a shell like the one to the right, which \emph{is} symmetric. In the right figure, where $\left\lvert\tfrac{1}{2}\vect{k}\right\rvert = \tfrac{1}{2}k$ is shown by the (horizontal) height of the orange-filled triangles towards the left, we see that as $k \to 0$ (the triangles shrink to their vertical baselines) the overlap increases and the resultant shifted shell is well approximated by the original, unshifted shell ($\cdot - \cdot$), as found in \eq{eq:shell_shift}.}
 \label{fig:shell_shift}
\end{figure}

We now turn our attention to the substitution $\vect{j} \rightarrow \tfrac{1}{2}\vect{k} + \vect{j}$ made by FNS and YO, under which our constraints become
\begin{equation}
 \theta^\high(\vect{j}) \theta^\high(\vect{k}-\vect{j}) \rightarrow \theta^\high(\tfrac{1}{2}\vect{k}+\vect{j}) \theta^\high(\tfrac{1}{2}\vect{k}-\vect{j})\ .
\end{equation}
Taylor expansion of these high-pass filters is now
\begin{eqnarray}
 \theta^\high(\tfrac{1}{2}\vect{k}\pm\vect{j}) &=& \theta^\high(\vect{j}) \pm \frac{\vect{k}\cdot\vect{j}}{2j}\Big[ \theta(\Lambda-j) \delta(j - D) - \theta(j-D) \delta(\Lambda-j) \Big] + {\mathcal O}(k^2) \nonumber \\
&=& \theta^\high(\vect{j}) \pm x(\vect{k},\vect{j}) + {\mathcal O}(k^2) \ ,
\end{eqnarray}
and the product becomes
\begin{equation}
 \label{eq:shell_shift}
 \theta^\high(\tfrac{1}{2}\vect{k}+\vect{j}) \theta^\high(\tfrac{1}{2}\vect{k}-\vect{j}) = \theta^\high(\vect{j}) + {\mathcal O}(k^2)\ .
\end{equation}
The contributions at ${\mathcal O}(k)$ cancel one another exactly, and there is no correction to the simple constraint on $\lvert\vect{j}\rvert$. Without the constraint $\theta^\high(\vect{k}-\vect{j})$, the substitution would have led 
to just $\theta^\high(\tfrac{1}{2}\vect{k}+\vect{j})$, which clearly does 
introduce a first order correction. These points can be seen in 
figure \ref{fig:shell_shift}. Using this, we go on to find the 
result of Yakhot \& Orszag (a generalization of the FNS result) in the 
limit $\omega \rightarrow 0$, $k \rightarrow 0$
\begin{eqnarray}
\label{eq:yo_result}
\Delta\nu_0 (\vect{0},0) &=& \frac{S_d}{(2\pi)^d} \frac{d^2 - d - \epsilon}{2d(d+2)} \frac{W_0 \lambda_0^2}{\nu_0^2 \Lambda^\epsilon} \left(\frac{e^{\epsilon\ell}-1}{\epsilon}\right) \nonumber \\
 &=& \nu_0 A_d \bar{\lambda}^2(0) \left(\frac{e^{\epsilon\ell} - 1}{\epsilon} \right) \ ,
\end{eqnarray}
where
\begin{equation}
 A_d = \frac{S_d}{(2\pi)^d} \tilde{A}_d\ , \quad\quad \tilde{A}_d = \frac{d^2 - d -\epsilon}{2d(d+2)}\ , \quad\quad \bar{\lambda}^2(0) = \frac{W_0 \lambda_0^2}{\nu_0^3 \Lambda^\epsilon} \ , \qquad \epsilon = 4 - d + y \ .
\end{equation}

\subsubsection{Wang \& Wu}
Wang \& Wu were unsatisfied with the substitution used by FNS and YO. This is because they do not impose the condition that $D < \lvert\vect{k}-\vect{j}\rvert < \Lambda$ on the self-energy integral, and on the face of things the substitution shifts the integration domain (see figure \ref{fig:shell_shift}). The authors then continue without making any substitution but simply Taylor expanding the denominator and expanding the vertex operator to ${\mathcal O}(k^2)$
\begin{eqnarray}
\label{eq:WW_current}
 \Sigma^\low_\alpha &=& \frac{4\lambda_0^2}{\nu_0^2} M_{\alpha\beta\gamma}(\vect{k}) u^\low_\mu(\vect{k},0) \int \frac{d^dj}{(2\pi)^d} \frac{W(j)}{j^2} M_{\gamma\mu\nu}(\vect{k}-\vect{j}) \frac{P_{\beta\nu}(\vect{j}) }{j^2 + \lvert\vect{k}-\vect{j}\rvert^2}\ \theta^{\high}(\vect{j}) \theta^{\high}(\vect{k}-\vect{j}) 
\nonumber \\
 &=& P^\low_{\alpha\beta\gamma\mu}(\vect{k}) \int d^dj\ j^{-y-4}\ \theta^{\high}(\vect{j}) \theta^{\high}(\vect{k}-\vect{j})
\nonumber \\
& & \left[k_\nu P_{\gamma\mu}(\vect{j})P_{\beta\nu}(\vect{j}) - k_\nu \frac{j_\mu j_\gamma}{j^2}P_{\beta\nu}(\vect{j}) - k_\nu \frac{j_\mu j_\nu}{j^2}P_{\beta\gamma}(\vect{j}) - j_\mu P_{\beta\gamma}(\vect{j})\right]\ ,
\end{eqnarray}
where the operator $P^\low_{\alpha\beta\gamma\mu}(\vect{k}) = \frac{W_0\lambda_0^2}{i\nu^2_0} \frac{M_{\alpha\beta\gamma}(\vect{k}) u^\low_\mu(\vect{k},0)}{(2\pi)^d}$ is defined for convenience. We now expand the product of $\theta^\high$ functions as in \eq{eq:ww_theta_exp} but note that only the last term in the square brackets above is not already of ${\mathcal O}(k)$. As such, it is the \emph{only} term that can generate a correction. 
$\Sigma^\low_\alpha$ then splits into
\begin{equation}
  \Sigma^\low_\alpha = \hat{\Sigma}^{\low}_\alpha + \delta \hat{\Sigma}^{\low}_\alpha \ ,
\end{equation}
where
\begin{equation}
 \hat{\Sigma}^{\low}_\alpha = P^\low_{\alpha\beta\gamma\mu}(\vect{k}) \int d^dj\ j^{-y-4}\ \theta^{\high}(\vect{j}) \left[k_\nu P_{\gamma\mu}(\vect{j})P_{\beta\nu}(\vect{j}) - k_\nu \frac{j_\mu j_\gamma}{j^2}P_{\beta\nu}(\vect{j}) - k_\nu \frac{j_\mu j_\nu}{j^2}P_{\beta\gamma}(\vect{j}) - j_\mu P_{\beta\gamma}(\vect{j})\right]
\end{equation}
is the contribution used by Wang \& Wu without imposing the additional constraint, and the correction
\begin{equation}
 \label{eq:ww_correction}
 \delta \hat{\Sigma}^{\low}_\alpha = P^\low_{\alpha\beta\gamma\mu}(\vect{k}) k_\nu \int d^dj\ j^{-y-4}\ \theta^\high(\vect{j})\frac{j_\mu j_\nu}{j} P_{\beta\gamma}(\vect{j})  \Big[ \theta(\Lambda-j) \delta(j - D) - \theta(j-D) \delta(\Lambda-j) \Big]
\end{equation}
includes the additional boundary terms.

The first contribution above leads to the Wang \& Wu result, that
\begin{equation}
 \hat{\Sigma}^{\low}_\alpha = -\Delta\nu^{_{WW}}_0 (\vect{k},0) k^2 u^\low_\alpha(\vect{k},0) \ ,
\end{equation}
where
\begin{equation}
 \label{eq:ww_result}
 \Delta\nu^{_{WW}}_0 (\vect{0},0) = \nu_0 A^\star_d \bar{\lambda}^2(0) \left(\frac{e^{\epsilon\ell} - 1}{\epsilon} \right)\ , \qquad A^\star_d = \frac{S_d}{(2\pi)^d} \tilde{A}^\star_d\ , \qquad \tilde{A}^\star_d = \frac{d-1}{2(d+2)} \ .
\end{equation}

We now evaluate the first order correction given by \eq{eq:ww_correction}, using the standard convention that $\theta(0) = \frac{1}{2}$  (see, for example, \cite{gozzi:1983-functional,hochberg:2000-effective_potential}),
\begin{eqnarray}
 \delta \hat{\Sigma}^{\low}_\alpha &=& P^\low_{\alpha\beta\gamma\mu}(\vect{k}) k_\nu \frac{S_d}{d(d+2)} \Big[(d+1)\delta_{\mu\nu} \delta_{\beta\gamma} - \delta_{\mu\beta} \delta_{\nu\gamma} - \delta_{\mu\gamma} \delta_{\nu\beta} \Big] \Big[D^{-\epsilon} - \Lambda^{-\epsilon}\Big] \theta(0) \nonumber \\
 &=& \frac{W_0\lambda_0^2}{\nu^2_0} \frac{S_d}{(2\pi)^d} \frac{1}{2d(d+2)} \left(\frac{e^{\epsilon\ell}-1}{\Lambda^{\epsilon}}\right) k^2 u^\low_\alpha(\vect{k},0) \ ,
\end{eqnarray}
and so the correction to the viscosity increment found by Wang \& Wu is
\begin{equation}\label{eq:ww_visc_correction}
 \delta\nu^{_{WW}}_0(\vect{k},0) = -\frac{W_0\lambda_0^2}{\nu^2_0} \frac{S_d}{(2\pi)^d} \frac{1}{2d(d+2)} \left(\frac{e^{\epsilon\ell}-1}{\Lambda^{\epsilon}}\right)\ .
\end{equation}
If this contribution is added to the result for the renormalized viscosity increment found by Wang \& Wu in \eq{eq:ww_result}, we find
\begin{equation}
\begin{split}
 \Delta\nu_0(\vect{0},0) &= \Delta\nu^{_{WW}}_0(\vect{0},0) + \delta\nu^{_{WW}}_0(\vect{0},0) \\
&= \frac{S_d}{(2\pi)^d} \left(\frac{d-1}{2(d+2)\epsilon} - \frac{1}{2d(d+2)}\right) \frac{W_0 \lambda_0^2}{\nu^2_0 \Lambda^\epsilon} \big(e^{\epsilon\ell} - 1\big) \\
&= \frac{S_d}{(2\pi)^d} \frac{d^2 - d - \epsilon}{2d(d+2)} \frac{W_0 \lambda_0^2}{\nu^2_0 \Lambda^\epsilon} \left(\frac{e^{\epsilon\ell} - 1}{\epsilon}\right) \ ,
\end{split}
\end{equation}
which is exactly the result obtained by YO, see \eq{eq:yo_result}. Hence we have shown that a more careful consideration of the region of integration used by Wang \& Wu instead leads to the result found by FNS and then later
YO. The approach taken by Teodorovich \cite{Teodorovich:1994-YO_theory} 
uses a different method for evaluating the angular part of the self-energy 
integral. However, the author misses the same constraint
and thus arrives at the result as Wang \& Wu.

\subsubsection{Nandy}\label{subsec:nandy}
In the paper by Nandy \cite{Nandy:1997-symm}, the author presents an argument based on symmetrising the self-energy integral. Referring to \eq{eq:all_agree}, 
he points out that there is no reason to do the $d^dp$ integral first, and 
that the result should be an average of the two. Performing the $d^dj$ 
integrals first, gives
\begin{equation}
 \Sigma^\low_\alpha(\vect{k},0) = \frac{4\lambda_0^2}{\nu_0^2} M_{\alpha\beta\gamma}(\vect{k}) u^\low_\mu(\vect{k},0) \int \frac{d^dp}{(2\pi)^d} \frac{W(\lvert\vect{k}-\vect{p}\rvert)}{\lvert\vect{k}-\vect{p}\rvert^2} M_{\gamma\mu\nu}(\vect{p}) \frac{P_{\beta\nu}(\vect{k}-\vect{p}) }{\lvert\vect{k}-\vect{p}\rvert^2 + p^2}\ \theta^{\high}(\vect{k}-\vect{p}) \theta^{\high}(\vect{p})  \ .
\end{equation}
Taylor expanding the function $\lvert\vect{k}-\vect{p}\rvert^{-y-2}$ and the 
denominator, and using the definition and properties of the vertex and 
projection operators, to ${\mathcal O}(k^2)$ leads to
\begin{equation}
 \begin{split}
  \Sigma^\low_\alpha(\fvect{k}) &= P^\low_{\alpha\beta\gamma\mu}(\vect{k}) \int d^dp\ p^{-y-4}\ \theta^{\high}(\vect{k}-\vect{p})\theta^{\high}(\vect{p}) \\
  &\qquad\qquad\times \left[ k_\nu P_{\gamma\nu}(\vect{p}) \frac{p_\mu p_\beta}{p^2} - k_\nu P_{\gamma\mu}(\vect{p}) \frac{p_\beta p_\nu}{p^2} + k_\beta P_{\gamma\mu}(\vect{p}) + (y+3)k_\nu P_{\beta\gamma}(\vect{p})\frac{p_\mu p_\nu}{p^2} + p_\mu P_{\beta\gamma}(\vect{p}) \right] \ .
 \end{split}
\end{equation}
Again, we see that all the terms in the square brackets are already ${\mathcal O}(k)$ except the last one, and so this is the only term which generates a correction. Once again decomposing
\begin{equation}
  \Sigma^\low_\alpha = \bar{\Sigma}^{\low}_\alpha + \delta \bar{\Sigma}^{\low}_\alpha \ ,
\end{equation}
the contribution calculated by Nandy is
\begin{equation}
 \begin{split}
  \bar{\Sigma}^{\low}_\alpha &= P^\low_{\alpha\beta\gamma\mu}(\vect{k}) \int d^dp\ p^{-y-4}\ \theta^{\high}(\vect{p}) \\
 &\qquad\qquad\times \left[ k_\nu P_{\gamma\nu}(\vect{p}) \frac{p_\mu p_\beta}{p^2} - k_\nu P_{\gamma\mu}(\vect{p}) \frac{p_\beta p_\nu}{p^2} + k_\beta P_{\gamma\mu}(\vect{p}) + (y+3)k_\nu P_{\beta\gamma}(\vect{p})\frac{p_\mu p_\nu}{p^2} + p_\mu P_{\beta\gamma}(\vect{p}) \right] \ ,
 \end{split}
\end{equation}
and the correction generated by expanding the product of $\theta^\high$ functions is given by
\begin{equation}
  \delta \bar{\Sigma}^{\low}_\alpha = -P^\low_{\alpha\beta\gamma\mu}(\vect{k}) k_\nu \int d^dp\ p^{-y-4}\ \theta^\high(\vect{p})\frac{p_\mu p_\nu}{p} P_{\beta\gamma}(\vect{p})  \Big[ \theta(\Lambda-p) \delta(p - D) - \theta(p-D) \delta(\Lambda-p) \Big] = - \delta \hat{\Sigma}^{\low}_\alpha \ .
\end{equation}
We see that, with the relabelling $\vect{p} \rightarrow \vect{j}$, the correction is exactly the same as \eq{eq:ww_correction} only with the opposite sign. Therefore, we see that
\begin{equation}
 \tfrac{1}{2}\big[ <\textrm{Wang \& Wu}> + <\textrm{Nandy}> \big] = \tfrac{1}{2}\big[ \hat{\Sigma}^{\low}_\alpha + \bar{\Sigma}^{\low}_\alpha \big] = \tfrac{1}{2}\big[ (\hat{\Sigma}^{\low}_\alpha + \delta \hat{\Sigma}^{\low}_\alpha) + (\bar{\Sigma}^{\low}_\alpha + \delta \bar{\Sigma}^{\low}_\alpha) \big] = \Sigma^{\low}_\alpha \ ,
\end{equation}
which is why this symmetrization produced the correct result. In fact, evaluation of $\bar{\Sigma}^{\low}_\alpha$ leads to the result found by Nandy for performing the integrals in this order
\begin{equation}
 \Delta\nu^{_N}_0(\vect{0},0) = \frac{W_0 \lambda^2}{\nu^2_0} \frac{S_d}{(2\pi)^d} \frac{d^2 - d -2\epsilon}{2d(d+2)} \left(\frac{e^{\epsilon\ell} - 1}{\epsilon\Lambda^\epsilon} \right) \ , \qquad d^2 - d -2\epsilon = d^2 + d - 2y - 8\ ,
\end{equation}
and the correction is
\begin{equation}\label{eq:nandy_visc_correction}
 \delta\nu^{_{N}}_0(\vect{0},0) = +\frac{W_0\lambda^2}{\nu^2_0} \frac{S_d}{(2\pi)^d} \frac{\epsilon}{2d(d+2)} \left(\frac{e^{\epsilon\ell}-1}{\epsilon\Lambda^{\epsilon}}\right)\ .
\end{equation}
Combining these results we again find the result of 
Yakhot \& Orszag, \eq{eq:yo_result}, showing that regardless of which 
integral is performed first we obtain the same result and so a 
symmetrization is not necessary.

In a completely different approach, Sukoriansky \textit{et al.} \cite{Sukoriansky_Galperin_Staroselsky:2003-cross_term} used a self-substitution method to solve for the low-frequency modes and claim to evaluate the cross-term exactly. This method does not generate the cubic linearity, instead it creates a contribution of the same form as Wang \& Wu \eq{eq:WW_current} but with the condition $\theta^\low(\vect{k}-\vect{j})$. Combined, this then covers the whole domain, and the authors drop any conditions on $\lvert\vect{k}-\vect{j}\rvert$. However,
\begin{equation}
 \theta^\low(\vect{k}-\vect{j}) + \theta^\high(\vect{k}-\vect{j}) = \theta(\Lambda - \lvert\vect{k}-\vect{j}\rvert) \neq 1 \ ,
\end{equation}
due to the upper momentum cutoff. This can then be Taylor expanded for small $k$, and leads to a contribution from the upper boundary, neglected in their analysis. In fact, this correction places their result somewhat between that found by Yakhot \& Orszag and Wang \& Wu,
\begin{equation}
 \begin{split}
 \Delta\nu_0(\vect{0},0) &= \frac{W_0 \lambda^2}{\nu^2_0} \left[A_d^\star  \left(\frac{(\Lambda e^{-\ell})^{-\epsilon} - \Lambda^{-\epsilon}}{\epsilon}\right) + \frac{\Lambda^{-\epsilon}}{2d(d+2)}\right] \\
 &= \frac{W_0 \lambda^2}{\nu^2_0} \left[ \frac{A_d^\star(\Lambda e^{-\ell})^{-\epsilon} - A_d(\epsilon)\Lambda^{-\epsilon}}{\epsilon}\right] \ ,
\end{split}
\end{equation}
where it is missing the contribution from the lower boundary. However, this self-substitution is not the same as solving a dynamical equation for the low-frequency modes and substituting for the high-frequency components. This method is fundamentally different from the standard RG procedure, and its result agreeing with neither YO or WW is further evidence that it is another approximation entirely.

\section{Noise Renormalization}\label{sec:renorm_cond}

\subsection{FNS treatment}\label{app:sub:FNS_conds}
In the paper by FNS \cite{FNS:1977-full}, the authors use two different scaling conditions (see appendix \ref{app:operator_scaling}) when analysing their models due to the contribution of the induced force to the renormalization. We first mention the results used by Yakhot \& Orszag for comparison (see appendix \ref{app:operator_scaling}),
\begin{equation}
 \chi = \tfrac{1}{2}(3z + d + y) \ , \qquad \chi = d+1 \qquad \longrightarrow \qquad z = 2 - \tfrac{\epsilon}{3}\ ,\quad \bar{\lambda}^{*2} = \frac{\epsilon}{3 A_d}\ ,
\end{equation}
where $\bar{\lambda}^{*2}$ is the reduced coupling at the non-trivial fixed point.

{\bfseries FNS model A} ($y = -2$):
For this model the authors show using diagrams (see figure 1 of \cite{FNS:1977-full}) how the propagator, force autocorrelation (shown here in figure \ref{fig:w_graph}) and vertex are renormalized. They conclude that $\nu_0$ and $W_0$ are renormalized the same way in their equations (3.10--11). This condition implies fixing the mean dissipation rate rather than $W_0$, and is then enforced under rescaling (see appendix \ref{app:operator_scaling}) by choosing $\chi = z + \frac{d}{2}$, which does not agree with $\chi$ above (first relation) used by YO with $y=-2$.

At this point, FNS invoke Galilean invariance (GI) to impose the condition that the vertex is not renormalized, such that $\lambda = \lambda_0 = 1$ to all orders in perturbation theory (in the limit of small external momenta, appendix B of \cite{FNS:1977-full}). While this is the case at 
$\vect{k} = \vect{0}$ \cite{B_H:2007-gauge_symm}, 
and as such does not invalidate the FNS theory, in general the consequences 
of the symmetry are trivial and do not lead to a condition on the vertex
\cite{B_H:2007-gauge_symm,McComb:2005-gi,B_H:2005-gi,B_H:2009-gauge_fix}. 
Further discussion is given in the conclusions, Sec. \ref{sec:conclusion}.
Taking the condition $\chi = d+1$ to preserve Galilean invariance, they find the non-trivial stable fixed point $\bar{\lambda}^{*2} = \epsilon/2 A_d$ (when $\epsilon > 0$).

{\bfseries FNS model B} ($y = 0$):
In this case, the one-loop graph in figure \ref{fig:w_graph} is claimed to be 
${\mathcal O}(k^2)$ and so cannot contribute to the constant part of the force 
autocorrelation. This term is then irrelevant and the force is rescaled 
accordingly.
This requires $\chi = \frac{1}{2}(3z + d)$, which is the same condition found 
by YO with $y=0$. Ensuring that Galilean invariance is satisfied, they have $\bar{\lambda}^{*2} = \epsilon/3 A_d$, as do YO.

This difference in scaling conditions leads to different differential equations and hence different solutions for the reduced coupling $\bar{\lambda}$ and the viscosity, depending on whether the noise is allowed to be renormalized or not. In the field-theoretic approach by Ronis \cite{Ronis:1987-field_theor_RG}, the force is also allowed to be renormalized, and the author comments that YO ignore this in their analysis. In fact, they restricted their work to $y > -2$ to avoid this issue.

This discrepancy only really applies to $y=-2$ when the noise coefficient is renormalized, although could lead to complications for $y < -2$ as the induced force always contributes as ${\mathcal O}(k^2)$ to the autocorrelation and becomes the leading order as $k \rightarrow 0$. In their paper \cite{YO:1986-RG_full}, YO state that ``in the limit $k \rightarrow 0$ this [induced] force is negligible in comparison with original forcing with $y<-2$'', and present an argument for neglecting it as equation (3.13). For the case $y > -2$ it is sub-leading and thus safely neglected. 
%--- @Arjun: Should this sentence stay?
This highlights another potential problem with calculating inertial range statistics, since it is only sub-leading as $k \to 0$.
%---

\subsection{Re-evaluation}\label{app:sub:re-eval}
\begin{figure}[tb]
 \begin{minipage}[t]{\columnwidth}
  \centering
  \includegraphics[width=0.55\columnwidth,bb=216 592 671 721]{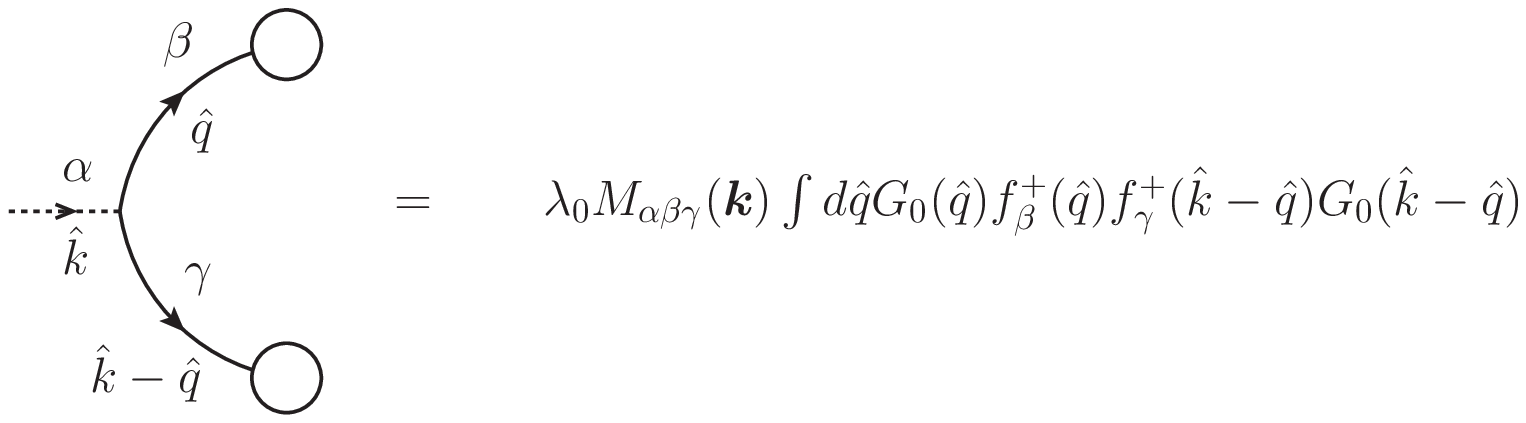}
  \caption{Graphical representation of the form of the induced random force given in \eq{eq:induced_force}. Note that the two momenta $\fvect{q}$, $\fvect{k}-\fvect{q}$ lie in the eliminated shell.}
  \label{fig:induced}
 \end{minipage}
\end{figure}

\begin{figure}[tb]
 \begin{minipage}[t]{0.95\columnwidth}
  \centering
  \includegraphics[width=0.5\columnwidth,bb=75 610 575 720]{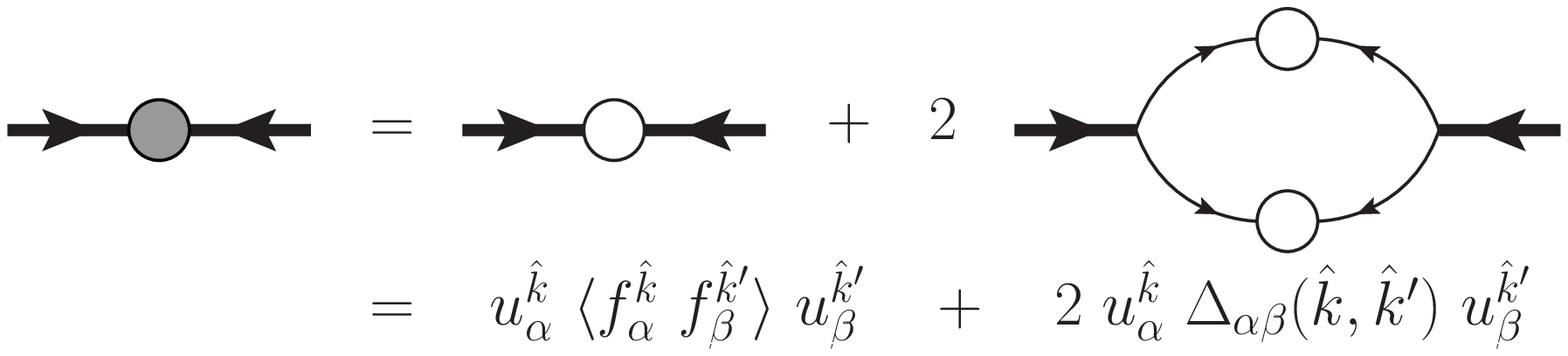}
  \caption{Renormalization of the force autocorrelation by FNS. See figure \ref{fig:rules} for Feynman rules.}
  \label{fig:w_graph}
 \end{minipage}
\end{figure}

In this section we will show how the induced force leads exactly to the 
graph in figure \ref{fig:w_graph}. We then evaluate the graph to analyse 
the contribution to the renormalization of the force.
For this, consider the form of the induced force shown in 
figure \ref{fig:induced}. Under averaging, we see that the graph forms a 
closed loop and is ${\mathcal O}(k)$ due to the vertex operator, and 
hence $\langle \Delta f_\alpha(\fvect{k})\rangle = 0$. The new random force
$\tilde{f}_\alpha^\low(\fvect{k}) = f_\alpha^\low(\fvect{k}) + \Delta f_\alpha^\low(\fvect{k})$ 
is invariant under the filtered-averaging procedure, and has autocorrelation
\begin{equation}
 \langle \tilde{f}_\alpha(\fvect{k}) \tilde{f}_\rho(\fvect{k}') \rangle = \langle f_\alpha(\fvect{k}) f_\rho(\fvect{k}') \rangle + \langle \Delta f_\alpha(\fvect{k}) \Delta f_\rho(\fvect{k}') \rangle\ .
\end{equation}
The new contribution due to the induced force is written as \eq{eq:app:delta_f_cor},
 \begin{equation}
  \label{eq:app:delta_f_cor}
  \begin{split}
   \langle \Delta f_\alpha(\fvect{k}) \Delta f_{\rho}(\fvect{k}') \rangle &= \lambda_0^2 M_{\alpha\beta\gamma}(\vect{k}) M_{\rho\mu\nu}(\vect{k}') \int\int \fmeasure{q}{\Omega}\fmeasure{p}{\Omega'} G_0(\fvect{q}) G_0(\fvect{k} - \fvect{q}) G_0(\fvect{p}) G_0(\fvect{k}' - \fvect{p}) \\
   &\qquad\qquad\qquad \times \langle f_\beta(\fvect{q}) f_\gamma(\fvect{k}-\fvect{q}) f_\mu(\fvect{p}) f_\nu(\fvect{k}'-\fvect{p}) \rangle \theta^\high(\vect{q}) \theta^\high(\vect{p}) \theta^\high(\vect{k}-\vect{q})  \theta^\high(\vect{k}'-\vect{p}) \ .
  \end{split}
 \end{equation}
Since the forcing is taken to be Gaussian, we may split the fourth-order moment
\begin{equation}
 \label{eq:split_4th}
  \langle f_\beta^{\fvect{q}} f_\gamma^{\fvect{k}-\fvect{q}} f_\mu^{\fvect{p}} f_\nu^{\fvect{k}'-\fvect{p}} \rangle =
  \langle f_\beta^{\fvect{q}} f_\gamma^{\fvect{k}-\fvect{q}} \rangle \langle f_\mu^{\fvect{p}} f_\nu^{\fvect{k}'-\fvect{p}} \rangle + \langle f_\beta^{\fvect{q}} f_\mu^{\fvect{p}} \rangle \langle f_\gamma^{\fvect{k}-\fvect{q}} f_\nu^{\fvect{k}'-\fvect{p}} \rangle 
  + \langle f_\beta^{\fvect{q}} f_\nu^{\fvect{k}'-\fvect{p}} \rangle \langle f_\mu^{\fvect{p}} f_\gamma^{\fvect{k}-\fvect{q}} \rangle \ ,
\end{equation}
and, using the definition of the force correlation \eq{eq:def:stirring_forces}, we see that the first contribution leads to two disconnected loops (which do not contribute to the force renormalization), whereas the other two both generate graphs like figure \ref{fig:w_graph} and appear to contribute towards the renormalization of $W_0$.


Using the rules given in figure \ref{fig:rules}, we write an analytic form for the ${\mathcal O}(\lambda_0^2)$ diagram in figure \ref{fig:w_graph} as \eq{eq:graph_analytic},
 \begin{equation}
  \label{eq:graph_analytic}
\begin{split}
  \Delta_{\alpha\rho}(\fvect{k},\fvect{k}') = 2 \lambda_0^2 M_{\alpha\beta\gamma}(\vect{k}) M_{\rho\mu\nu}(\vect{k}') \int\int \fmeasure{q}{\Omega}\fmeasure{p}{\Omega'} G_0(\fvect{q}) \langle f_\beta(\fvect{q}) f_\mu(\fvect{p}) \rangle G_0(\fvect{p}) \\
G_0(\fvect{k} - \fvect{q}) \langle f_\gamma(\fvect{k}-\fvect{q}) f_\nu(\fvect{k}'-\fvect{p}) \rangle G_0(\fvect{k}' - \fvect{p}) \ .
\end{split}
 \end{equation}
The factor $2$ is due to symmetry of the graph (figure 2 in \cite{Wyld:1961-formulation}). This may be compared to the correlation of the induced force given by \eq{eq:app:delta_f_cor} which, along with the requirement that momenta of all internal lines in \eq{eq:graph_analytic} are in the eliminated shell, agree exactly.

An outline of the evaluation of this correction to leading order is given in appendix \ref{sec:app:eval_noise}. As a result of our analysis, we find
\begin{equation}
 \label{eq:noise_result}
 \langle \tilde{f}_\alpha(\fvect{k}) \tilde{f}_{\rho}(\fvect{k}') \rangle = 2 W_0 k^{-y} P_{\alpha\rho}(\vect{k}) \delta(\fvect{k}+\fvect{k}') \left[1 + \bar{\lambda}^2(0) B_d  \left(\frac{e^{\ell(\epsilon+y+2)} - 1}{(\epsilon+y+2)\Lambda_0^{y+2}}\right) k^{y+2} \right] \ ,
\end{equation}
where
\begin{equation}
 B_d = \frac{S_d}{(2\pi)^d} \frac{d^2 - 2}{2d(d+2)}\ .
\end{equation}

We see for the equilibrium case $y = -2$ that the correction may be taken as an multiplicative renormalization to $W_0$ and write
\begin{equation}
 \begin{split}
  &\langle \tilde{f}_\alpha(\fvect{k}) \tilde{f}_{\rho}(\fvect{k}') \rangle = 2 W_I k^2 P_{\alpha\rho}(\vect{k}) \delta(\fvect{k}+\fvect{k}')\ ,
 \end{split}
\end{equation}
with
\begin{equation}
 W_I = W_0 \left[1 + \bar{\lambda}^2(0) B_d  \left(\frac{e^{\epsilon\ell} - 1}{\epsilon}\right) \right]\ .
\end{equation}
This may be compared to equation (3.11) of FNS \cite{FNS:1977-full} and (3.4b) of Ronis \cite{Ronis:1987-field_theor_RG} (with $y = -2$). We note that both authors find the noise coefficient and viscosity to be renormalized with the same prefactor, what we have defined as $B_d$. However, this prefactor only coincides with $A_d$ found by FNS and YO for $y = -2$ ($\epsilon = 2-d$), and the analysis is only valid for the equilibrium case (the work by Ronis is an expansion about $y=-2$).

The prefactor $B_d$ was calculated to leading order with no change of variables in a similar fashion to Wang \& Wu, and agrees with $A_d$ found by FNS and that by YO with $y=-2$. If the prefactor $A_d^\star$ found by Wang \& Wu and others, which is $\epsilon$-independent, were the true expression, we should have recovered it from this analysis also. Instead, it only agrees with $B_d$ when $d = 2$ (the critical dimension for $y=-2$ (where $\epsilon = 0$ and $A_d^\star = A_d$)). Taking the induced contribution as an multiplicative renormalization only when $y=-2$, $A_d = B_d$ whereas only $A^\star_{d=2} = B_{d=2}$. We feel this supports our argument that the correct expression for the prefactor is the $\epsilon$-dependent result found by YO.

As noted by Ronis, the pole structure leading to logarithmic divergence 
in the noise renormalization more generally occurs when $\epsilon+y+2= 0$.
However for the case $y = -2$,  we recover the same pole in $\epsilon$ 
found by FNS and in the viscosity renormalization. Since noise renormalization is only meaningful in this case, the presentation of equation (3.4b) in \cite{Ronis:1987-field_theor_RG} as a general result seems misleading.

In summary, we have calculated the renormalized noise coefficient to one-loop and find the prefactor $B_d$ to agree with FNS as well as YO when
setting $y=-2$ in their result. While noise renormalization was not considered by Wang \& Wu, we have found their $\epsilon$-free result, $A_d^\star$, to only agree with $B_d$ in 2-dimensions. For $y > -2$, this induced forced correlation becomes sub-leading and is ignored, as assumed in the YO analysis. When $y < -2$, the induced contribution does not renormalize the noise coefficient but will be the leading term as $k \rightarrow 0$. In this case, it is not clear how to interpret the validity of the results obtained, since the forcing appears on large scales to be dominated by the order $\bar{\lambda}^2(0)$ contribution, making the viscosity calculation order $\bar{\lambda}^4(0)$, \textit{i.e.} two-loop, which has not been done here.

\section{Conclusion}\label{sec:conclusion}
\renewcommand\arraystretch{2}
\tablinesep = 2ex
\begin{table}
 \begin{tabular}{l|c|c|c|c|c|c}
 & Our analysis & FNS {\bfseries A} & FNS {\bfseries B} & YO & WW / T / N & Ronis \\
\hline
 Viscosity & $\tilde{A}_d(\epsilon)$ & $\tilde{A}_d(\epsilon)\big\rvert_{y=-2}$ & $\tilde{A}_d(\epsilon)\big\rvert_{y=0}$ & $\tilde{A}_d(\epsilon)$ & $\tilde{A}_d^\star = \tilde{A}_d(0)$ & $\tilde{A}_d(\epsilon)\big\rvert_{y=-2}$\\
 \hline
 Noise & $\tilde{B}_d = \tilde{A}_d(\epsilon)\big\rvert_{y=-2}$ & $\tilde{B}_d$ & --- & $\tilde{B}_d$ & --- & $\tilde{B}_d$ \\
 \hline
 Pole structure & $1/(\epsilon + y + 2)$ & $1/\epsilon$ & --- & $1/\epsilon$ & --- & $1/(\epsilon + y + 2)$
\end{tabular}
\caption{A summary of the prefactors for the viscosity increment, noise renormalization, and the $\epsilon$-pole  structure found in the various analyses considered in this paper. These expressions are valid for all $y$, with the exception of FNS models A ($y=-2$) and B ($y=0$). `T' represents Teodorovich and `N' Nandy. $\tilde{A}_d(\epsilon) = {(d^2 - d - \epsilon)}/{2d(d+2)}$ and hence $\tilde{B}_d = {(d^2 - 2)}/{2d(d+2)}$. By pole structure we mean the $\epsilon$-dependence of the denominator for the induced noise correlation (see equations (\ref{eq:yo_result}) and (\ref{eq:noise_result}), for example). Our analysis agrees with FNS and YO. The work of Ronis appears to be for general $y$, but the viscosity is only in agreement for $y=-2$.}
\label{tbl:summary}
\end{table}

A summary of our results and a comparison with other authors
is presented in table \ref{tbl:summary}. We conclude that the 
analysis of FNS does not suffer from a shifted domain of integration 
in the self-energy integral which is evaluated due to the constraint 
$\lvert\vect{k}-\vect{j}\rvert < \Lambda$, neglected by other authors. 
Using $\theta$ functions to control the integration domain, we have shown 
that the corrections cancel exactly at first order in $k$ when the change 
of variables is made. We then showed that this ignored constraint leads to 
a correction in the Wang \& Wu- and Nandy-style calculations which exactly 
reproduces the result found by 
YO. The noise renormalization for the case $y=-2$ was then shown, 
using a substitution-free method similar to Wang \& Wu, to lead to a prefactor 
compatible with YO for all $d$ and only compatible with Wang \& Wu for 
$d=2$, which we feel supports our claim as to the validity of the 
FNS and YO results.

That said, some comments should be made on the application of this method to 
calculating inertial range statistics, which may not be so well justified.
Despite its applicability only on the largest of length scales, Yakhot \& Orszag use the expressions obtained with this infra-red procedure to calculate inertial range properties \cite{YO:1986-RG_full}, such as the Kolmogorov constant. To do this, they use a set of assumptions that they term the \emph{correspondence principle}.

Briefly, the correspondence principle states that an unforced system which 
started from some initial conditions with a developed inertial range is 
statistically equivalent to a system forced in such a way as to generate the 
same scaling exponents. 
In particular if forcing is introduced to generate the scaling exponents
at low $k$, this artificially 
generated ``inertial range'' can then be used to calculate values 
for various inertial range parameters using the properties of universality. 
There is an implicit assumption that, as long as the scaling exponents match,
all other quantities will also match. This may be the reason that YO raise the 
cutoff $\Lambda$ out of the production range to ${\mathcal O}(k_d)$ 
(see above equation (2.2) in \cite{YO:1986-RG_full}) so that the 
renormalization passes through the inertial range, whereas FNS explicitly 
consider $\Lambda \ll k_d$ (final paragraph of their section II.A).

YO find that when $y=d$ the noise coefficient, $W_0$, has the dimensions of the dissipation rate, $\varepsilon$, and they take $W_0 = a\varepsilon$ (with $a$ constant). They can then obtain a Kolmogorov scaling region when $\epsilon = 4$ is used, but also require $\epsilon = 0$ in the prefactor $A_d(\epsilon)$ in the same equation. This has been unsatisfactory for many authors, and appears to favour the $\epsilon$-free result found by Wang \& Wu, as then $\epsilon = 4$ alone reproduces the famous $k^{-5/3}$ result. However, we have shown why the $\epsilon$-free result is incorrect.

There are still a number of technical difficulties associated with taking $\epsilon \to 4$ and generating a $k^{-5/3}$ spectrum:
\begin{itemize}
 \item {The Wilson-style $\epsilon$ expansion is valid only for $\epsilon$ small, and there is no evidence that results will be valid at $\epsilon = 4$. The neglected cubic and higher-order non-linear terms generated by iterating this procedure may not be irrelevant, and there is no estimate of the accumulation of error even for $\epsilon \sim 0$, let alone $\epsilon \rightarrow 4$. In the review by Smith \& Woodruff \cite{smith_woodruff:1998-rg_anal_turb}, they discuss the only justification for the validity of $\epsilon \rightarrow 4$ being that it leads to good agreement with inertial range constants, and describe it as ``intriguing and difficult to interpret''. They also present an argument for YO's use of $\epsilon = 0$ in the prefactor, it being required for a self-consistent asymptotic expansion at each iteration step.}
 \item {The IR behaviour as $k \to 0$ is dominated by the fixed point which, for $\epsilon > 0$, is at $\bar{\lambda}^{\ast} = (\epsilon/3A_d)^{1/2}$. To lowest order in $\epsilon$, this is then evaluated with $A_d(\epsilon = 0)$. However, $\epsilon$ is no longer small, nor is $\bar{\lambda}^{\ast}$. In 3-dimensions with $\epsilon = 4$, this fixed point is at $\bar{\lambda}^{\ast} = (20\pi^2\epsilon/(6-\epsilon))^{1/2} \simeq 11.5$ to leading order in $\epsilon$, or $\simeq 20$ when evaluated to all orders.}
 \item {As shown by figure \ref{fig:energy_spectrum}, the asymptotic nature of this renormalization scheme taking us to the infra-red means we don't enter the inertial range, and are always sensitive to the forcing spectrum.}
 \item {The forcing spectrum required to obtain $k^{-5/3}$ is divergent as $k\rightarrow 0$ ($\epsilon = 4$ requires $y=d>1$, so $W(k) \sim k^{-d}$), as is the energy spectrum itself. As shown by McComb \cite{mccomb:1990-book}, ensuring that there is a balance between energy input at large length scales and energy dissipated at small (this is statistically stationary turbulence) we see that the range of forced wavenumbers predicted by their analysis has $k_t/k_b \simeq 1.007$, where $k_t$ and $k_b$ are, respectively, the upper and lower bounding wavenumbers of the input range. The energy input is also logarithmically divergent as $k_t \rightarrow \infty$ or $k_b \rightarrow 0$.}
 \item {The condition of Galilean invariance (GI) used by FNS and adopted 
by YO to enforce the non-renormalization of the vertex at all orders is 
actually only valid at $k = 0$ \cite{B_H:2007-gauge_symm}. 
In general, the consequences of GI are 
trivial and provide no constraint on the vertex 
\cite{McComb:2005-gi,B_H:2007-gauge_symm,B_H:2009-gauge_fix,B_H:2005-gi}. This is supported by recent numerical results \cite{wio:2010-KPZ} from a KPZ model on a discretized lattice with a broken GI symmetry, which have found the same critical exponents as the actual KPZ model (which does possess GI)  \cite{FNS:1977-full,MHKZ:1989-correlated}, even though GI has been explicitly violated. This questions the connection between GI and the scaling relations associated
with the critical exponents.
% The scaling relations used to ensure non-renormalization of the vertex which were attributed to GI (see appendix \ref{app:operator_scaling}) are therefore only valid when $k=0$.Recent numerical results \cite{wio:2010-KPZ} of the KPZ model on a discretized lattice with a broken GI symmetry have found the same critical exponents, even though GI has been explicitly violated, as those which do possess this invariance \cite{FNS:1977-full,MHKZ:1989-correlated}. This shows that GI is not responsible for constraining the vertex.
As such, care must be taken when extending this theory to $k \neq 0$. 
This introduces another issue for the study of inertial range properties 
using the correspondence principle, as $k$ cannot be chosen to lie in the 
inertial range without the vertex being renormalized.}
 \item {The assumed Gaussian lowest-order behaviour of the fluid is only valid 
at the smallest wavenumbers when subject to Gaussian forcing, since the 
response of the system is then also Gaussian. However, this assumption 
cannot be translated to the inertial range, which should be insensitive 
to the details of the energy input and is inherently 
non-Gaussian \cite{McComb:2006-ngpt}.
% It simply receives energy and passes down the line.
 }
\end{itemize}
The need to use two different values for $\epsilon$ in the same formula to estimate inertial range properties is therefore not the only failure of this scheme.

The solution for the renormalized viscosity at the largest scales ($\ell \rightarrow \infty$) can be found to behave as
\begin{equation}
 \nu(\mu) \sim \left(\frac{3 A_d}{\epsilon}\right)^{1/3} W_0^{1/3} \mu^{-\epsilon/3} \ ,
\end{equation}
where $\mu = \Lambda e^{-\ell} \rightarrow 0$ is the new cut off. 
With the assumption $W_0 = a\varepsilon$, interestingly this does have 
the same form as that found by other methods (e.g. \cite{McComb:2006-ngpt}), 
that the viscosity is proportional to $\varepsilon^{1/3}$ and, 
with $\epsilon = 4$, the cut off $\mu^{-4/3}$. However, there is an important 
difference: the cut off $\mu$ is going to zero for this expression to hold, 
which is not the location of the inertial range, unlike the iterative 
averaging approach by McComb \cite{McComb:1982-reform,McComb:1992-two_field}. 
Smith \& Woodruff note \cite{smith_woodruff:1998-rg_anal_turb} that this is 
not dependent on the dissipation range quantities $\Lambda$ and $\nu_0$, as 
inertial range coefficients should be. But it is still dependent on the 
forcing spectrum through $\epsilon$, which it should not be.

\section*{Acknowledgements}
We thank David McComb for suggesting this
project and for his helpful advice during it.
We were funded by STFC.

\appendix

\section{Operator scaling}\label{app:operator_scaling}
Although rescaling the variables after performing an iteration of the renormalization procedure outlined above is not performed in the calculation of the renormalized viscosity (and thus it can be argued not to be an RG procedure \cite{Eyink:1994-RG_method_stat_hyd}), it is still useful to consider how the rescaling would affect the equations of motion. Using a scaling factor, $s$, the spatial coordinates transform as $\vect{x} = s\vect{x}'$ and $t = s^z t'$ (where the unprimed variables are the original scale), with $s > 1$, and so $\vect{k} = s^{-1}\vect{k}'$, $\omega = s^{-z}\omega'$ with $\vect{u}(\vect{k},\omega) = s^\chi \vect{u}'(\vect{k}',\omega')$. In YO \cite{YO:1986-RG_full}, $s = e^\ell$, $\alpha(\ell) = z\ell$ and $\zeta(\ell) = e^{\chi\ell} = s^\chi$. Equation (\ref{eq:def:nse_fourier}) then transforms under the scaling to
\begin{equation}
\begin{split}
 \left(i\omega' + (s^{z-2}\nu) k'^2\right) u'_\alpha(\vect{k}',\omega') = (s^{z-\chi})f_\alpha(\vect{k},\omega) + (s^{\chi - (d+1)}\lambda) M_{\alpha\beta\gamma}(\vect{k}') \int \measure{j'}{\Omega'} \\
u_\beta(\vect{j}',\Omega') u_\gamma(\vect{k}'-\vect{j}',\omega'-\Omega')\ ,
\end{split}
\end{equation}
and so we find
\begin{eqnarray}
 \label{eq:nu'}
 \nu' = \nu(s) &=& s^{z-2}\nu\ , \\
 \label{eq:lambda'}
 \lambda' = \lambda(s) &=& s^{\chi - (d+1)}\lambda\ ,
\end{eqnarray}
\begin{equation}
 \label{eq:force'}
 \begin{split}
 f'_\alpha(\vect{k}',\omega') &= s^{z-\chi} f_\alpha(\vect{k},\omega) \\
 &= s^{z-\chi + \frac{1}{2}(y+d+z)} f_\alpha(\vect{k}',\omega')\ .
 \end{split}
\end{equation}
Using \eq{eq:force'} with the definition of the force autocorrelations \eq{eq:def:stirring_forces} and the scaling for $\vect{k}$ and $\omega$, we find
\begin{equation}
 \label{eq:W_0'}
 W' = W(s) = s^{3z - 2\chi + d + y} W\ .
\end{equation}
Equations (\ref{eq:nu'})--(\ref{eq:W_0'}) agree with (2.28)--(2.33) of \cite{YO:1986-RG_full}. Due to Galilean invariance, as $k \rightarrow 0$ \eq{eq:lambda'} is forced to give the condition that $\chi = d+1$. For $y\neq -2$, the elimination of scales should not affect $W_0$ as there is no multiplicative renormalization, a condition that must also be preserved under scaling to find $\chi = \frac{1}{2}\left(3z + d + y\right)$. YO note (from (2.34) of \cite{YO:1986-RG_full}) that the renormalized viscosity \emph{at the fixed point} is $s$-independent if $z = 2 - A_d \bar{\lambda}^{*2} = 2 - \frac{\epsilon}{3}$.

As noted by FNS and discussed in section \ref{sec:renorm_cond}, for the case $y = -2$ we must consider the renormalization of $W_0$ and instead require that $\nu$ and $W$ be renormalized in the same way, \textit{i.e.} $\chi = z +\frac{d}{2}$. This scaling condition is not the same as that for $y \neq -2$, and leads to a different solution.

Under the YO prescription, the triple non-linearity 
\begin{equation}
\mu M_{\alpha\beta\gamma}(\fvect{k}) \int\fmeasure{j}{\Omega} \int\fmeasure{p}{\Omega'} M_{\gamma\mu\nu}(\fvect{k}-\fvect{j}) G_0(\fvect{k}-\fvect{j}) u^{\low}_\beta(\fvect{j}) u^{\low}_\mu(\fvect{p}) u^{\low}_\nu(\fvect{k}-\fvect{j}-\fvect{p})
\end{equation}
gives, using the expression $\chi = \frac{1}{2}\left(3z + d + y\right)$,
\begin{equation}
 \label{eq:trip_nonlin}
 \begin{split}
 \mu' = \mu(s) \sim s^{2(\chi - d - 1)} \mu &= s^{3z - 2 + (y - d)} \mu \\
&= s^{3(z - (2-\frac{\epsilon}{3}))} \mu\ .
 \end{split}
\end{equation}
This should be compared to equation (2.45) in \cite{YO:1986-RG_full}, which reads
\begin{equation}
 \mu' \sim s^{2\chi - (2d+2)} \mu = s^{-(d-y)} \mu\ .
\end{equation}
They comment that for $y < d$ ($\epsilon < 4$) the operator is irrelevant, and marginal when $y = d$ ($\epsilon = 4$). However, we see that their result requires $\chi = \frac{1}{2}(d+y+2)$, which only agrees with the above expression for $\chi$ (ensuring $W_0$ is not altered) when $z = \frac{2}{3} = \left.(2 - \frac{\epsilon}{3})\right\vert_{\epsilon = 4}$. Therefore, they have already used $\epsilon = 4$ to obtain this result. If we do not specify $\epsilon$ but do require that $z = 2 - \frac{\epsilon}{3}$ (so that the viscosity at the fixed point is $s$-independent), we see from \eq{eq:trip_nonlin} that $\mu' \sim \mu$ and the operator is not irrelevant but marginal. (This could also have been seen by requiring $\chi = d+1$ in \eq{eq:trip_nonlin}, and we see that if the vertex is not renormalized the triple moment cannot be irrelevant.) This is discussed in a paper by Eyink \cite{Eyink:1994-RG_method_stat_hyd}. Attempts to retain the effects of the triple non-linearity on the viscosity increment are analysed in \cite{ZVH:1988-RG_theory, carati:1991-locality, smith:1991-exploration}.

\section{Taylor expansion of $\theta^\high$-functions}\label{app:theta_exp}
We here describe the procedure for Taylor expanding a $\theta^\high$-function.
The high-band filter $\theta^\high$ is defined as
\begin{equation}
 \theta^\high(\vect{j}) = \theta(\lvert\vect{j}\rvert-A) \theta(\Lambda-\lvert\vect{j}\rvert)\ , \qquad A=\Lambda e^{-\ell} < \Lambda\ ,
\end{equation}
where the first restricts us to $j > A$ and the second to $j < \Lambda$. The $\theta$-function product of consideration here is
\begin{equation}
 \theta^\high(\vect{j}) \theta^\high(\vect{k}-\vect{j})\ ,
\end{equation}
and we Taylor expand $\theta^\high(\vect{k}-\vect{j}) = \theta(\lvert\vect{k}-\vect{j}\rvert-A) \theta(\Lambda-\lvert\vect{k}-\vect{j}\rvert)$ as:
\begin{eqnarray}
 \theta(\lvert\vect{k}-\vect{j}\rvert-A) &=& \theta(\lvert\vect{k}-\vect{j}\rvert-A)\Big\rvert_{\vect{k}=0} + \vect{k}\cdot\Big(\vect{\nabla}\theta(\lvert\vect{k}-\vect{j}\rvert-A)\Big)\Big\rvert_{\vect{k}=0} + \ldots \nonumber \\
&=& \theta(\lvert\vect{j}\rvert-A) + \vect{k}\cdot\left(\frac{\vect{k}-\vect{j}}{\lvert\vect{k}-\vect{j}\rvert}\right) \Big(\delta(\lvert\vect{k}-\vect{j}\rvert-A)\Big)\Big\rvert_{\vect{k}=0} + \ldots \nonumber \\
&=& \theta(\lvert\vect{j}\rvert-A) - \frac{\vect{k}\cdot\vect{j}}{\lvert\vect{j}\rvert} \delta(\lvert\vect{j}\rvert-A) + {\mathcal O}(k^2)
\ ,
\end{eqnarray}

\begin{eqnarray}
 \theta(\Lambda-\lvert\vect{k}-\vect{j}\rvert) &=& \theta(\Lambda-\lvert\vect{k}-\vect{j}\rvert)\Big\rvert_{\vect{k}=0} + \vect{k}\cdot\Big(\vect{\nabla}\theta(\Lambda-\lvert\vect{k}-\vect{j}\rvert)\Big)\Big\rvert_{\vect{k}=0} + \ldots 
\nonumber \\
&=& \theta(\Lambda-\lvert\vect{j}\rvert) - \vect{k}\cdot\left(\frac{\vect{k}-\vect{j}}{\lvert\vect{k}-\vect{j}\rvert}\right) \Big(\delta(\Lambda-\lvert\vect{k}-\vect{j}\rvert)\Big)\Big\rvert_{\vect{k}=0} + \ldots \nonumber \\
&=& \theta(\Lambda-\lvert\vect{j}\rvert) + \frac{\vect{k}\cdot\vect{j}}{\lvert\vect{j}\rvert} \delta(\Lambda-\lvert\vect{j}\rvert) + {\mathcal O}(k^2)
\ .
\end{eqnarray}
Our expansion is then
\begin{equation}
 \theta^\high(\vect{k}-\vect{j}) = \theta^\high(\vect{j}) - \frac{\vect{k}\cdot\vect{j}}{j}\Big[ \theta(\Lambda-j) \delta(j - A) - \theta(j-A) \delta(\Lambda-j) \Big] + {\mathcal O}(k^2)\ .
\end{equation}

\section{Evaluation of the noise renormalization}\label{sec:app:eval_noise}
We now evaluate the correlation of the induced force to leading order using a more compact notation. Starting from equations (\ref{eq:app:delta_f_cor}--\ref{eq:split_4th}),

 \begin{equation}
  \label{eq:app:f_calc}
\begin{split}
   \langle \Delta f_\alpha^{\fvect{k}} \Delta f_{\rho}^{\fvect{k}'} \rangle = \lambda_0^2 M_{\alpha\beta\gamma}^{\vect{k}} M_{\rho\mu\nu}^{\vect{k}'} \int\int \fmeasure{q}{\Omega}\fmeasure{p}{\Omega'} G_0^{\fvect{q}} G_0^{\fvect{k} - \fvect{q}} G_0^{\fvect{p}} G_0^{\fvect{k}' - \fvect{p}} 
   \left[ \langle f_\beta^{\fvect{q}} f_\mu^{\fvect{p}} \rangle \langle f_\gamma^{\fvect{k}-\fvect{q}} f_\nu^{\fvect{k}'-\fvect{p}} \rangle + \langle f_\beta^{\fvect{q}} f_\nu^{\fvect{k}'-\fvect{p}} \rangle \langle f_\mu^{\fvect{p}} f_\gamma^{\fvect{k}-\fvect{q}} \rangle \right] \\
\theta^\high_{\vect{q}} \theta^\high_{\vect{p}} \theta^\high_{\vect{k}-\vect{q}}  \theta^\high_{\vect{k}'-\vect{p}} \ ,
\end{split}
\end{equation}
we note that our integrals here are unconstrained and the shell of integration is controlled by the $\theta^\high$-functions. Since the substitution $\fvect{p} \rightarrow \fvect{p}' = \fvect{k}'-\fvect{p}$ preserves the product $\theta^\high({\vect{p}}) \theta^\high({\vect{k}'-\vect{p}}) \rightarrow \theta^\high({\vect{p}'}) \theta^\high({\vect{k}'-\vect{p}'}) = \theta^\high({\vect{k}'-\vect{p}}) \theta^\high({\vect{p}})$, we may use it, along with the property of the vertex operator $M_{\rho\mu\nu}(\vect{k}') = M_{\rho\nu\mu}(\vect{k}')$ and index relabelling for the second term in the square brackets, to combine the two contributions and write
 \begin{equation}
   \langle \Delta f_\alpha^{\fvect{k}} \Delta f_{\rho}^{\fvect{k}'} \rangle = 2\lambda_0^2 M_{\alpha\beta\gamma}^{\vect{k}} M_{\rho\mu\nu}^{\vect{k}'} \int\int \fmeasure{q}{\Omega}\fmeasure{p}{\Omega'} G_0^{\fvect{q}} G_0^{\fvect{k} - \fvect{q}} G_0^{\fvect{p}} G_0^{\fvect{k}' - \fvect{p}} 
   \langle f_\beta^{\fvect{q}} f_\mu^{\fvect{p}} \rangle \langle f_\gamma^{\fvect{k}-\fvect{q}} f_\nu^{\fvect{k}'-\fvect{p}} \rangle \theta^\high_{\vect{q}} \theta^\high_{\vect{p}} \theta^\high_{\vect{k}-\vect{q}}  \theta^\high_{\vect{k}'-\vect{p}} \ ,
\end{equation}
which we see is exactly \eq{eq:graph_analytic}. This reveals that the symmetry factor of 2 associated to the graph is due to exchanging legs on the vertex. In to this we substitute the definition of the force autocorrelation, then integration over $\fvect{p}$ is trivially done using the $\delta$-functions obtained to give
\begin{equation}
\langle \Delta f_\alpha^{\fvect{k}} \Delta f_{\rho}^{\fvect{k}'} \rangle = 8\lambda_0^2 M_{\alpha\beta\gamma}^{\vect{k}} M_{\rho\mu\nu}^{\vect{k}'} \delta(\fvect{k}+\fvect{k}') \int \fmeasure{q}{\Omega} \left\lvert G_0^{\fvect{q}} \right\rvert^2 \left\lvert G_0^{\fvect{k}-\fvect{q}} \right\rvert^2 W(q) W(\lvert\vect{k}-\vect{q}\rvert) P_{\beta\mu}^{\vect{q}} P_{\gamma\nu}^{\vect{k}-\vect{q}} \theta^\high_{\vect{q}} \theta^\high_{\vect{k}-\vect{q}}\ .
\end{equation}
The constraint enforced by the remaining $\delta$-function is then used to restrict $\fvect{k}' = -\fvect{k}$, along with the property $M_{\rho\mu\nu}(\vect{k}') \delta(\vect{k}+\vect{k}') = - M_{\rho\mu\nu}(\vect{k}) \delta(\vect{k}+\vect{k}')$, resulting in
\begin{equation}
 \label{eq:app:noise_1}
 \langle \Delta f_\alpha^{\fvect{k}} \Delta f_{\rho}^{\fvect{k}'} \rangle = -8\lambda_0^2 M_{\alpha\beta\gamma}^{\vect{k}} M_{\rho\mu\nu}^{\vect{k}} \delta(\fvect{k}+\fvect{k}') \int \fmeasure{q}{\Omega} \left\lvert G_0^{\fvect{q}} \right\rvert^2 \left\lvert G_0^{\fvect{k}-\fvect{q}} \right\rvert^2 W(q) W(\lvert\vect{k}-\vect{q}\rvert) P_{\beta\mu}^{\vect{q}} P_{\gamma\nu}^{\vect{k}-\vect{q}} \theta^\high_{\vect{q}} \theta^\high_{\vect{k}-\vect{q}} \ .
\end{equation}
The frequency integral is then performed, closing the contour in the upper-halfplane and collecting the residue from two poles $\Omega = i\nu_0 q^2$ and $\Omega = \omega + i\nu_0\lvert\vect{k}-\vect{q}\rvert$, with the result
\begin{equation}
 \begin{split}
  \int d\Omega\ \left\lvert G_0^{\fvect{q}} \right\rvert^2 \left\lvert G_0^{\fvect{k}-\fvect{q}} \right\rvert^2 =& \frac{\pi}{\nu_0}\left[\frac{q^2\left(i\omega + \nu_0(q^2 + \lvert\vect{k}-\vect{q}\rvert^2)\right) + \lvert\vect{k}-\vect{q}\rvert^2\left(i\omega - \nu_0(q^2 + \lvert\vect{k}-\vect{q}\rvert^2)\right)}{q^2 \lvert\vect{k}-\vect{q}\rvert^2 \left(\omega^2 + \nu_0^2(q^2 + \lvert\vect{k}-\vect{q}\rvert^2)^2 \right)}\right] \\
&\qquad \times \left[\frac{1}{i\omega + \nu_0(q^2 - \lvert\vect{k}-\vect{q}\rvert^2)}\right] \\
  \stackrel{\omega\rightarrow 0}{=}& \frac{\pi}{\nu_0^3}\left[\frac{1}{q^2 \lvert\vect{k}-\vect{q}\rvert^2 \left(q^2 + \lvert\vect{k}-\vect{q}\rvert^2 \right)}\right] \ .
 \end{split}
\end{equation}
The limit $\omega\rightarrow 0$ offers a huge simplification to the result. This is inserted in to \eq{eq:app:noise_1}
\begin{equation}
 \langle \Delta f_\alpha^{\vect{k}} \Delta f_{\rho}^{\vect{k}'} \rangle = \frac{-4\lambda_0^2 W_0^2}{\nu_0^3} M_{\alpha\beta\gamma}^{\vect{k}} M_{\rho\mu\nu}^{\vect{k}} \delta(\fvect{k}+\fvect{k}') \int \frac{d^dq}{(2\pi)^d} \left(q\lvert\vect{k}-\vect{q}\rvert\right)^{-y-2} \frac{P_{\beta\mu}^{\vect{q}} P_{\gamma\nu}^{\vect{k}-\vect{q}}}{q^2 + \lvert\vect{k}-\vect{q}\rvert^2} \theta^\high_{\vect{q}} \theta^\high_{\vect{k}-\vect{q}} \ ,
\end{equation}
and the integrand is expanded to leading order in $k$ as $k\rightarrow 0$
\begin{equation}
\langle \Delta f_\alpha^{\vect{k}} \Delta f_{\rho}^{\vect{k}'} \rangle = \frac{-2\lambda_0^2 W_0^2}{\nu_0^3} M_{\alpha\beta\gamma}^{\vect{k}} M_{\rho\mu\nu}^{\vect{k}} \delta(\fvect{k}+\fvect{k}') \int \frac{d^dq}{(2\pi)^d} q^{-2(y+3)} P_{\beta\mu}^{\vect{q}} P_{\gamma\nu}^{\vect{q}} \theta^\high_{\vect{q}} + {\mathcal O}(k^3) \ .
\end{equation}
Note that there is a power of $k$ associated to each of the vertex operators, hence the leading contribution will \emph{always} go as $k^2$. Expanding the function $\theta^\high(\vect{k}-\vect{q})$ we do not generate corrections as we are working to zero-order in $k$ in the integrand and the corrections are ${\mathcal O}(k)$. Expanding the projection operators and performing the $(d-1)$ angular integrals we find
\begin{equation}
\begin{split}
 \langle \Delta f_\alpha^{\vect{k}} \Delta f_{\rho}^{\vect{k}'} \rangle = \frac{-2\lambda_0^2 W_0^2}{\nu_0^3 (2\pi)^d} M_{\alpha\beta\gamma}^{\vect{k}} M_{\rho\mu\nu}^{\vect{k}} \delta(\fvect{k}+\fvect{k}') \frac{S_d}{d(d+2)} \left[ (d^2 - 3) \delta_{\beta\mu}\delta_{\gamma\nu} + \delta_{\beta\gamma}\delta_{\mu\nu} + \delta_{\beta\nu}\delta_{\gamma\mu} \right] \\
\int dq\ q^{-2(y+3)+d-1} \theta^\high_{\vect{q}}\ ,
\end{split}
\end{equation}
where we expand the vertex operators, do the remaining integral and perform contractions to obtain
\begin{equation}
 \label{eq:app:noise_2}
 \begin{split}
  \langle \Delta f_\alpha(\vect{k},0) \Delta f_{\rho}(\vect{k}',0) \rangle &= \frac{\lambda_0^2 W_0^2}{\nu_0^3 (2\pi)^d} \delta(\fvect{k}+\fvect{k}') \frac{S_d}{2d(d+2)} \left[ 2k^2 P_{\alpha\rho}(\vect{k}) (d^2 - 2) \right] \left(\frac{e^{\ell(\epsilon+y+2)} - 1}{(\epsilon+y+2)\Lambda_0^{\epsilon+y+2}}\right) \ ,
 \end{split}
\end{equation}
which we rearrange to our final result
\begin{equation}
\langle \Delta f_\alpha(\vect{k},0) \Delta f_{\rho}(\vect{k}',0) \rangle = 2W_0 \bar{\lambda}^2(0) B_d  \left(\frac{e^{\ell(\epsilon+y+2)} - 1}{(\epsilon+y+2)\Lambda_0^{y+2}}\right) k^2 P_{\alpha\rho}(\vect{k}) \delta(\fvect{k}+\fvect{k}')\ , \quad B_d = \frac{S_d}{(2\pi)^d} \frac{d^2 - 2}{2d(d+2)}\ .
\end{equation}

%--- Bibliography inserted manually
%\bibliographystyle{apsrev}
%\bibliography{../bib/papers,../bib/books,../bib/thesis}

\begin{thebibliography}{32}
%--- Is this needed?
\expandafter\ifx\csname natexlab\endcsname\relax\def\natexlab#1{#1}\fi
\expandafter\ifx\csname bibnamefont\endcsname\relax
  \def\bibnamefont#1{#1}\fi
\expandafter\ifx\csname bibfnamefont\endcsname\relax
  \def\bibfnamefont#1{#1}\fi
\expandafter\ifx\csname citenamefont\endcsname\relax
  \def\citenamefont#1{#1}\fi
\expandafter\ifx\csname url\endcsname\relax
  \def\url#1{\texttt{#1}}\fi
\expandafter\ifx\csname urlprefix\endcsname\relax\def\urlprefix{URL }\fi
\providecommand{\bibinfo}[2]{#2}
\providecommand{\eprint}[2][]{\url{#2}}
%--- ??


\bibitem[{\citenamefont{Forster et~al.}(1976)\citenamefont{Forster, Nelson, and
  Stephen}}]{FNS:1976-letter}
\bibinfo{author}{\bibfnamefont{D.}~\bibnamefont{Forster}},
  \bibinfo{author}{\bibfnamefont{D.~R.} \bibnamefont{Nelson}},
  \bibnamefont{and} \bibinfo{author}{\bibfnamefont{M.~J.}
  \bibnamefont{Stephen}}, \bibinfo{journal}{Phys. Rev. Lett.}
  \textbf{\bibinfo{volume}{36}}, \bibinfo{pages}{867} (\bibinfo{year}{1976}).

\bibitem[{\citenamefont{Forster et~al.}(1977)\citenamefont{Forster, Nelson, and
  Stephen}}]{FNS:1977-full}
\bibinfo{author}{\bibfnamefont{D.}~\bibnamefont{Forster}},
  \bibinfo{author}{\bibfnamefont{D.~R.} \bibnamefont{Nelson}},
  \bibnamefont{and} \bibinfo{author}{\bibfnamefont{M.~J.}
  \bibnamefont{Stephen}}, \bibinfo{journal}{Phys. Rev. A}
  \textbf{\bibinfo{volume}{16}}, \bibinfo{pages}{732} (\bibinfo{year}{1977}).

\bibitem[{\citenamefont{McComb}(2006)}]{McComb:2006-ngpt}
\bibinfo{author}{\bibfnamefont{W.~D.} \bibnamefont{McComb}},
  \bibinfo{journal}{Phys. Rev. E} \textbf{\bibinfo{volume}{73}},
  \bibinfo{pages}{026303} (\bibinfo{year}{2006}).

\bibitem[{\citenamefont{Yakhot and
  Orszag}(1986{\natexlab{a}})}]{YO:1986-RG_letter}
\bibinfo{author}{\bibfnamefont{V.}~\bibnamefont{Yakhot}} \bibnamefont{and}
  \bibinfo{author}{\bibfnamefont{S.~A.} \bibnamefont{Orszag}},
  \bibinfo{journal}{Phys. Rev. Lett.} \textbf{\bibinfo{volume}{57}},
  \bibinfo{pages}{1722} (\bibinfo{year}{1986}{\natexlab{a}}).

\bibitem[{\citenamefont{Yakhot and
  Orszag}(1986{\natexlab{b}})}]{YO:1986-RG_full}
\bibinfo{author}{\bibfnamefont{V.}~\bibnamefont{Yakhot}} \bibnamefont{and}
  \bibinfo{author}{\bibfnamefont{S.~A.} \bibnamefont{Orszag}},
  \bibinfo{journal}{J. Sci. Comp.} \textbf{\bibinfo{volume}{1}},
  \bibinfo{pages}{3} (\bibinfo{year}{1986}{\natexlab{b}}).

\bibitem[{\citenamefont{Teodorovich}(1994)}]{Teodorovich:1994-YO_theory}
\bibinfo{author}{\bibfnamefont{{\'{E}}.~V.} \bibnamefont{Teodorovich}},
  \bibinfo{journal}{Fluid Dynamics} \textbf{\bibinfo{volume}{29}},
  \bibinfo{pages}{770} (\bibinfo{year}{1994}).

\bibitem[{\citenamefont{McComb}(1990)}]{mccomb:1990-book}
\bibinfo{author}{\bibfnamefont{W.~D.} \bibnamefont{McComb}},
  \emph{\bibinfo{title}{The Physics of Fluid Turbulence}}
  (\bibinfo{publisher}{Oxford University Press}, \bibinfo{year}{1990}).

\bibitem[{\citenamefont{Eyink}(1994)}]{Eyink:1994-RG_method_stat_hyd}
\bibinfo{author}{\bibfnamefont{G.~L.} \bibnamefont{Eyink}},
  \bibinfo{journal}{Phys. Fluids} \textbf{\bibinfo{volume}{6}},
  \bibinfo{pages}{3063} (\bibinfo{year}{1994}).

\bibitem[{\citenamefont{Wang and Wu}(1993)}]{Wang_Wu:1993-mod_YO}
\bibinfo{author}{\bibfnamefont{X.-H.} \bibnamefont{Wang}} \bibnamefont{and}
  \bibinfo{author}{\bibfnamefont{F.}~\bibnamefont{Wu}}, \bibinfo{journal}{Phys.
  Rev. E} \textbf{\bibinfo{volume}{48}}, \bibinfo{pages}{R37}
  (\bibinfo{year}{1993}).

\bibitem[{\citenamefont{Adzhemyan et~al.}(1999)\citenamefont{Adzhemyan,
  Antonov, and Vasiliev}}]{AAV:1999-book}
\bibinfo{author}{\bibfnamefont{L.~T.} \bibnamefont{Adzhemyan}},
  \bibinfo{author}{\bibfnamefont{N.~V.} \bibnamefont{Antonov}},
  \bibnamefont{and} \bibinfo{author}{\bibfnamefont{A.~N.}
  \bibnamefont{Vasiliev}}, \emph{\bibinfo{title}{The Field Theoretic
  Renormalization Group in Fully Developed Turbulence}}
  (\bibinfo{publisher}{Gordon and Breach}, \bibinfo{year}{1999}),
  \bibinfo{note}{translated from the Russian by P. Millard.}

\bibitem[{\citenamefont{Nandy}(1997)}]{Nandy:1997-symm}
\bibinfo{author}{\bibfnamefont{M.~K.} \bibnamefont{Nandy}},
  \bibinfo{journal}{Phys. Rev. E} \textbf{\bibinfo{volume}{55}},
  \bibinfo{pages}{5455} (\bibinfo{year}{1997}).

\bibitem[{\citenamefont{Kardar et~al.}(1986)\citenamefont{Kardar, Parisi, and
  Zhang}}]{KPZ:1986-scaling}
\bibinfo{author}{\bibfnamefont{M.}~\bibnamefont{Kardar}},
  \bibinfo{author}{\bibfnamefont{G.}~\bibnamefont{Parisi}}, \bibnamefont{and}
  \bibinfo{author}{\bibfnamefont{Y.-C.} \bibnamefont{Zhang}},
  \bibinfo{journal}{Phys. Rev. Lett..} \textbf{\bibinfo{volume}{56}},
  \bibinfo{pages}{889} (\bibinfo{year}{1986}).

\bibitem[{\citenamefont{Burgers}(1948)}]{Burgers:1948-eq}
\bibinfo{author}{\bibfnamefont{J.~M.} \bibnamefont{Burgers}},
  \bibinfo{journal}{Adv. Appl. Mech.} \textbf{\bibinfo{volume}{1}},
  \bibinfo{pages}{171} (\bibinfo{year}{1948}).

\bibitem[{\citenamefont{Medina et~al.}(1989)\citenamefont{Medina, Hwa, Kardar,
  and Zhang}}]{MHKZ:1989-correlated}
\bibinfo{author}{\bibfnamefont{E.}~\bibnamefont{Medina}},
  \bibinfo{author}{\bibfnamefont{T.}~\bibnamefont{Hwa}},
  \bibinfo{author}{\bibfnamefont{M.}~\bibnamefont{Kardar}}, \bibnamefont{and}
  \bibinfo{author}{\bibfnamefont{Y.-C.} \bibnamefont{Zhang}},
  \bibinfo{journal}{Phys. Rev. A} \textbf{\bibinfo{volume}{39}},
  \bibinfo{pages}{3053} (\bibinfo{year}{1989}).

\bibitem[{\citenamefont{Frey and T\"{a}uber}(1994)}]{Frey_Tauber:1994-two-loop}
\bibinfo{author}{\bibfnamefont{E.}~\bibnamefont{Frey}} \bibnamefont{and}
  \bibinfo{author}{\bibfnamefont{U.~C.} \bibnamefont{T\"{a}uber}},
  \bibinfo{journal}{Phys. Rev. E} \textbf{\bibinfo{volume}{50}},
  \bibinfo{pages}{1024} (\bibinfo{year}{1994}).

\bibitem[{\citenamefont{Fournier et~al.}(1982)\citenamefont{Fournier, Sulem,
  and Pouquet}}]{FSP:1982-ir_mhd}
\bibinfo{author}{\bibfnamefont{J.-D.} \bibnamefont{Fournier}},
  \bibinfo{author}{\bibfnamefont{P.-L.} \bibnamefont{Sulem}}, \bibnamefont{and}
  \bibinfo{author}{\bibfnamefont{A.}~\bibnamefont{Pouquet}},
  \bibinfo{journal}{J. Phys. A} \textbf{\bibinfo{volume}{15}},
  \bibinfo{pages}{1393} (\bibinfo{year}{1982}).

\bibitem[{\citenamefont{Dannevik et~al.}(1987)\citenamefont{Dannevik, Yakhot,
  and Orszag}}]{YOD:1987-analytic_theories}
\bibinfo{author}{\bibfnamefont{W.~P.} \bibnamefont{Dannevik}},
  \bibinfo{author}{\bibfnamefont{V.}~\bibnamefont{Yakhot}}, \bibnamefont{and}
  \bibinfo{author}{\bibfnamefont{S.~A.} \bibnamefont{Orszag}},
  \bibinfo{journal}{Phys. Fluids} \textbf{\bibinfo{volume}{30}},
  \bibinfo{pages}{2021} (\bibinfo{year}{1987}).

\bibitem[{\citenamefont{Ronis}(1987)}]{Ronis:1987-field_theor_RG}
\bibinfo{author}{\bibfnamefont{D.}~\bibnamefont{Ronis}},
  \bibinfo{journal}{Phys. Rev. A} \textbf{\bibinfo{volume}{36}},
  \bibinfo{pages}{3322} (\bibinfo{year}{1987}).

\bibitem[{\citenamefont{McComb}(1982)}]{McComb:1982-reform}
\bibinfo{author}{\bibfnamefont{W.~D.} \bibnamefont{McComb}},
  \bibinfo{journal}{Phys. Rev. A} \textbf{\bibinfo{volume}{26}},
  \bibinfo{pages}{1078} (\bibinfo{year}{1982}).

\bibitem[{\citenamefont{McComb and Watt}(1990)}]{McComb:1990-ca}
\bibinfo{author}{\bibfnamefont{W.~D.} \bibnamefont{McComb}},
  \bibinfo{author}{\bibfnamefont{A.~G.} \bibnamefont{Watt}},
  \bibinfo{journal}{Phys. Rev. Lett..} \textbf{\bibinfo{volume}{65}},
  \bibinfo{pages}{3281} (\bibinfo{year}{1990}).

\bibitem[{\citenamefont{McComb and Watt}(1992)}]{McComb:1992-two_field}
\bibinfo{author}{\bibfnamefont{W.~D.} \bibnamefont{McComb}} \bibnamefont{and}
  \bibinfo{author}{\bibfnamefont{A.~G.} \bibnamefont{Watt}},
  \bibinfo{journal}{Phys. Rev. A} \textbf{\bibinfo{volume}{46}},
  \bibinfo{pages}{4797} (\bibinfo{year}{1992}).

\bibitem[{\citenamefont{Hunter}(2002)}]{thesis:ahunter}
\bibinfo{author}{\bibfnamefont{A.}~\bibnamefont{Hunter}}, Ph.D. thesis,
  \bibinfo{school}{University of Edinburgh} (\bibinfo{year}{2002}).

\bibitem[{\citenamefont{McComb et~al.}(1992)}]{McComb:1992-ca}
\bibinfo{author}{\bibfnamefont{W.~D.} \bibnamefont{McComb}},
  \bibinfo{author}{\bibfnamefont{W.} \bibnamefont{Roberts}}, \bibnamefont{and}
  \bibinfo{author}{\bibfnamefont{A.~G.} \bibnamefont{Watt}},
  \bibinfo{journal}{Phys. Rev. A} \textbf{\bibinfo{volume}{45}},
  \bibinfo{pages}{3507} (\bibinfo{year}{1992}).

\bibitem[{\citenamefont{Sukoriansky et~al.}(2003)\citenamefont{Sukoriansky,
  Galperin, and
  Staroselsky}}]{Sukoriansky_Galperin_Staroselsky:2003-cross_term}
\bibinfo{author}{\bibfnamefont{S.}~\bibnamefont{Sukoriansky}},
  \bibinfo{author}{\bibfnamefont{B.}~\bibnamefont{Galperin}}, \bibnamefont{and}
  \bibinfo{author}{\bibfnamefont{I.}~\bibnamefont{Staroselsky}},
  \bibinfo{journal}{Fluid Dyn. Research} \textbf{\bibinfo{volume}{33}},
  \bibinfo{pages}{319} (\bibinfo{year}{2003}).

\bibitem[{\citenamefont{Zhou et~al.}(1997)\citenamefont{Zhou, McComb, and
  Vahala}}]{ZhouMcCombVahala:1997-RG_in_turb}
\bibinfo{author}{\bibfnamefont{Y.}~\bibnamefont{Zhou}},
  \bibinfo{author}{\bibfnamefont{W.~D.} \bibnamefont{McComb}},
  \bibnamefont{and} \bibinfo{author}{\bibfnamefont{G.}~\bibnamefont{Vahala}},
  \bibinfo{journal}{NASA Contractor Rep.} \textbf{\bibinfo{volume}{201718}}
  (\bibinfo{year}{1997}).

\bibitem[{\citenamefont{Wyld}(1961)}]{Wyld:1961-formulation}
\bibinfo{author}{\bibfnamefont{H.~W.} \bibnamefont{Wyld}},
  \bibinfo{journal}{Ann. Phys.} \textbf{\bibinfo{volume}{14}},
  \bibinfo{pages}{143} (\bibinfo{year}{1961}).

%%--- FROM REFEREE #2
\bibitem[{\citenamefont{Gozzi}(1983)}]{gozzi:1983-functional}
\bibinfo{author}{\bibfnamefont{E.}~\bibnamefont{Gozzi}},
  \bibinfo{journal}{Phys. Rev. D} \textbf{\bibinfo{volume}{28}},
  \bibinfo{pages}{1922} (\bibinfo{year}{1983}).

\bibitem[{\citenamefont{Hochberg et~al.}(2000)}]{hochberg:2000-effective_potential}
\bibinfo{author}{\bibfnamefont{D.} \bibnamefont{Hochberg}},
  \bibinfo{author}{\bibfnamefont{C.}~\bibnamefont{Molina-Par\'is}},
  \bibinfo{author}{\bibfnamefont{J.}~\bibnamefont{P\'erez-Mercader}}, \bibnamefont{and}
  \bibinfo{author}{\bibfnamefont{M.}~\bibnamefont{Visser}},
  \bibinfo{journal}{Physica A} \textbf{\bibinfo{volume}{280}},
  \bibinfo{pages}{437} (\bibinfo{year}{2000}).

%%--- END FROM REFEREE

\bibitem[{\citenamefont{Berera and Hochberg}(2007)}]{B_H:2007-gauge_symm}
\bibinfo{author}{\bibfnamefont{A.}~\bibnamefont{Berera}} \bibnamefont{and}
  \bibinfo{author}{\bibfnamefont{D.}~\bibnamefont{Hochberg}},
  \bibinfo{journal}{Phys. Rev. Lett..} \textbf{\bibinfo{volume}{99}},
  \bibinfo{pages}{254501} (\bibinfo{year}{2007}).

\bibitem[{\citenamefont{McComb}(2005)}]{McComb:2005-gi}
\bibinfo{author}{\bibfnamefont{W.~D.} \bibnamefont{McComb}},
  \bibinfo{journal}{Phys. Rev. E} \textbf{\bibinfo{volume}{71}},
  \bibinfo{pages}{037301} (\bibinfo{year}{2005}).

\bibitem[{\citenamefont{Berera and Hochberg}(2005)}]{B_H:2005-gi}
\bibinfo{author}{\bibfnamefont{A.}~\bibnamefont{Berera}} \bibnamefont{and}
  \bibinfo{author}{\bibfnamefont{D.}~\bibnamefont{Hochberg}},
  \bibinfo{journal}{Phys. Rev. E} \textbf{\bibinfo{volume}{72}},
  \bibinfo{pages}{057301} (\bibinfo{year}{2005}).

\bibitem[{\citenamefont{Berera and Hochberg}(2009)}]{B_H:2009-gauge_fix}
\bibinfo{author}{\bibfnamefont{A.}~\bibnamefont{Berera}} \bibnamefont{and}
  \bibinfo{author}{\bibfnamefont{D.}~\bibnamefont{Hochberg}},
  \bibinfo{journal}{Nucl. Phys. B} \textbf{\bibinfo{volume}{814}},
  \bibinfo{pages}{522} (\bibinfo{year}{2009}).

\bibitem[{\citenamefont{Smith and
  Woodruff}(1998)}]{smith_woodruff:1998-rg_anal_turb}
\bibinfo{author}{\bibfnamefont{L.~M.} \bibnamefont{Smith}} \bibnamefont{and}
  \bibinfo{author}{\bibfnamefont{S.~L.} \bibnamefont{Woodruff}},
  \bibinfo{journal}{Ann. Rev. Fluid Mech.} \textbf{\bibinfo{volume}{30}},
  \bibinfo{pages}{275} (\bibinfo{year}{1998}).

%%--- FROM REFEREE #2
\bibitem[{\citenamefont{Wio et~al.}(2010)}]{wio:2010-KPZ}
\bibinfo{author}{\bibfnamefont{H.~S.} \bibnamefont{Wio}},
  \bibinfo{author}{\bibfnamefont{J.~A.}~\bibnamefont{Revelli}},
  \bibinfo{author}{\bibfnamefont{R.~R.}~\bibnamefont{Deza}},
  \bibinfo{author}{\bibfnamefont{C.}~\bibnamefont{Escudero}}, \bibnamefont{and}
  \bibinfo{author}{\bibfnamefont{M.~S.}~\bibnamefont{de~La~Lama}},
  \bibinfo{journal}{Phys. Rev. E} \textbf{\bibinfo{volume}{81}},
  \bibinfo{pages}{066706} (\bibinfo{year}{2010}).
%%--- END FROM REFEREE

\bibitem[{\citenamefont{Zhou et~al.}(1988)\citenamefont{Zhou, Vahala, and
  Hossain}}]{ZVH:1988-RG_theory}
\bibinfo{author}{\bibfnamefont{Y.}~\bibnamefont{Zhou}},
  \bibinfo{author}{\bibfnamefont{G.}~\bibnamefont{Vahala}}, \bibnamefont{and}
  \bibinfo{author}{\bibfnamefont{M.}~\bibnamefont{Hossain}},
  \bibinfo{journal}{Phys. Rev. A} \textbf{\bibinfo{volume}{37}},
  \bibinfo{pages}{2590} (\bibinfo{year}{1988}).

\bibitem[{\citenamefont{Carati}(1991)}]{carati:1991-locality}
\bibinfo{author}{\bibfnamefont{D.}~\bibnamefont{Carati}},
  \bibinfo{journal}{Phys. Rev. A} \textbf{\bibinfo{volume}{44}},
  \bibinfo{pages}{6932} (\bibinfo{year}{1991}).

\bibitem[{\citenamefont{Smith et~al.}(1991)\citenamefont{Smith, Waleffe, and
  Carati}}]{smith:1991-exploration}
\bibinfo{author}{\bibfnamefont{L.~M.} \bibnamefont{Smith}},
  \bibinfo{author}{\bibfnamefont{F.}~\bibnamefont{Waleffe}}, \bibnamefont{and}
  \bibinfo{author}{\bibfnamefont{D.}~\bibnamefont{Carati}},
  \bibinfo{journal}{AFOSR Rep.} \textbf{\bibinfo{volume}{91}},
  \bibinfo{pages}{0272} (\bibinfo{year}{1991}).

\end{thebibliography}

\end{document}